\documentclass[a4paper,10pt]{article}

\usepackage{amssymb}
\usepackage{amsfonts}
\usepackage{amsmath}
\DeclareMathOperator\arctanh{arctanh}
\usepackage{mathtools}
\usepackage{latexsym}
\usepackage{graphicx,color}
\usepackage{caption}
\graphicspath{ {images/} }
\usepackage{subcaption}
\usepackage{mathrsfs}
\usepackage[colorlinks=true, allcolors=violet]{hyperref}
\usepackage{tikz}
\usepackage{mdwlist}
\usepackage{cite}

\newskip\humongous \humongous=0pt plus 1000pt minus 1000pt

\newif\ifdtup

\allowdisplaybreaks[1]

\catcode`\@=11

\@addtoreset{equation}{section}
\def\theequation{\thesection.\arabic{equation}}

\def\@normalsize{\@setsize\normalsize{15pt}\xiipt\@xiipt
\abovedisplayskip 14pt plus3pt minus3pt%
\belowdisplayskip \abovedisplayskip
\abovedisplayshortskip \z@ plus3pt%
\belowdisplayshortskip 7pt plus3.5pt minus0pt}

\def\small{\@setsize\small{13.6pt}\xipt\@xipt
\abovedisplayskip 13pt plus3pt minus3pt%
\belowdisplayskip \abovedisplayskip
\abovedisplayshortskip \z@ plus3pt%
\belowdisplayshortskip 7pt plus3.5pt minus0pt
\def\@listi{\parsep 4.5pt plus 2pt minus 1pt
     \itemsep \parsep
     \topsep 9pt plus 3pt minus 3pt}}

\relax

\catcode`@=12

\catcode`\@=11

\def\section{\@startsection{section}{1}{\z@}{3.5ex plus 1ex minus
   .2ex}{2.3ex plus .2ex}{\large\bf}}

\def\SymBoxes#1#2#3#4{\newdimen\un@t \un@t#3%
\raisebox{#1}{\rule{#2\un@t}{#4}\hskip-#2\un@t
\@tempdimb\un@t \advance\@tempdimb by-#4\@tempcntb#2\relax%
\@whilenum{\@tempcntb>0}\do{
\rule{#4}{\un@t}\hskip\@tempdimb \advance\@tempcntb by\m@ne}%
\hskip-#2\un@t \rule[\un@t]{#2\un@t}{#4}%
\rule[\un@t]{#4}{#4}\hskip-#4
\rule{#4}{\un@t}}\hskip-#4}                

\begin{document}

\newcommand{\beq}{\begin{equation}}
	\newcommand{\eeq}{\end{equation}}
\newcommand{\bea}{\begin{align}}
	\newcommand{\eea}{\end{align}}
\newcommand{\bos}{\boldsymbol}
\newcommand{\bom}{\boldsymbol{\omega}}
\newcommand{\boep}{\boldsymbol{\epsilon}}
\newcommand{\boj}{\bos{\mathit{j}}}
\newcommand{\hdel}{\hat{\delta}}
\newcommand{\dv}{\delta v}
\newcommand{\ed}{\mathrm{d}}
\newcommand{\tom}{\mathtt{\Omega}}
\def\cA{{\cal{A}}}
\def\cB{{\cal{B}}}
\def\cC{{\cal{C}}}
\def\cL{{\cal{L}}}
\def\cP{{\cal{P}}}
\def\cV{{\cal{V}}}
\def\cM{{\cal{M}}}
\def\m{\mu}
\def\n{\nu}
\def\p{\phi}
\def\P{\Phi}
\def\trho{\tilde{\rho}}
\def\pa{\partial}
\def\pr{\prime}

\numberwithin{equation}{section}

\newcommand{\pkd}[1]{\textcolor{red}{\textsf{[Pratik: #1]}}}
\newcommand{\mm}[1]{\textcolor{blue}{\textsf{[Manav: #1]}}}

\begin{titlepage}
\def\thefootnote{\fnsymbol{footnote}}


\begin{center}
{\large
{\bf Bit Threads: From Entanglement to Geometric Entropies
}
}
\end{center}

\begin{center}
{Pratik K. Das\footnote{\texttt{dask.pratik@gmail.com}}, \ \ Manavendra Mahato\footnote{\texttt{manav@iiti.ac.in} }}


\end{center}

\renewcommand{\thefootnote}{\arabic{footnote}}

\begin{center}
\vspace{-0.2cm}
 {\small \sl Department of Physics,} \\
{\small \sl Indian Institute of Technology Indore,} \\
{\small \sl Simrol, Khandwa Road, 453552, Indore, India}

\end{center}

\noindent
\begin{center} {\bf Abstract} \end{center}
In this work, we attempt to construct bit thread configurations for various backgrounds using expressions from the covariant phase space formalism. We find that when the Ryu-Takayanagi surface is same as the horizon, such expressions are sufficient. In other cases, it differs by gradient of a harmonic function. We explore its relation to Wald and differential entropy, and re-express the first law of entanglement entropy in terms of bit threads. Inclusion of quantum effects imposes some constraints on the bulk entanglement via the dominant energy condition. We also apply our method to ascertain a bit thread configuration in a certain dynamical spacetime.

\vspace{1.6 cm}

\vfill

\end{titlepage}

\setcounter{page}{2}
\tableofcontents

\setcounter{footnote}{0}


\section{Introduction}\label{sec:introduction}

Holographic entanglement entropy has emerged as a useful concept in gauge/gravity duality, providing a crucial bridge between quantum information theory and the geometry of spacetime \cite{Ryu:2006bv}. It takes us one step closer to understanding various issues regarding the quantum nature of gravity, such as the black hole information problem and many aspects of the AdS/CFT correspondence. The Ryu-Takayanagi (RT) formula equates the entanglement entropy $S_A$ defined in the dual conformal field theory (CFT) on a boundary region $A$ to the area of a minimal codimension-2 surface $m(A)$ (homologous to $A$) of the bulk asymptotically Anti-de Sitter (AdS) spacetime \cite{Ryu:2006bv, Ryu:2006ef}:
\beq
 S_A = \dfrac{Area(m(A))}{4 G_N}.
\eeq
 Recently, Freedman and Headrick \cite{Freedman:2016zud} introduced a novel, flow-based method for computing holographic entanglement entropy (HEE) that bypasses the need for prior identification of the RT surface. In this approach, the entanglement entropy is obtained by maximising the flux of a divergenceless vector field over the boundary region. These vector field lines, termed ``bit threads", can be interpreted as representing the quantum correlations (or ``Bell pairs") that connect different parts of the boundary \cite{Freedman:2016zud}. Bit threads thus provide a geometric representation of entanglement, offering an alternative perspective to the traditional minimal surface prescription. In this framework, the concept of Bell pairs plays a central role in interpreting the physical meaning of bit threads. Bell pairs, which represent maximally entangled two-level quantum systems, serve as the fundamental units of entanglement. In the bit thread picture, each thread can be thought of as carrying the information of one Bell pair. In other words, the number of bit threads that connect a given boundary region to its complement reflects the number of Bell pairs shared between these regions\footnote{It has been shown for pure bipartite states in the context of tensor networks in \cite{Mori:2022xec}.}, thereby quantifying the entanglement entropy. This interpretation is motivated by the idea that entanglement entropy is a measure of the nonlocal quantum correlations between subsystems. By associating each bit thread with a Bell pair, the flow-based prescription provides an intuitive geometrical representation of these correlations. The maximisation of bit thread flux, which is equivalent to finding the minimal surface in the traditional Ryu-Takayanagi prescription, effectively counts the number of Bell pairs across the boundary. While the identification of an individual thread with a specific Bell pair is a heuristic picture, it offers an alternate way to visualise the abstract notion of entanglement in holographic theories. This perspective not only reinforces the connection between geometry and quantum information but also opens avenues for exploring how modifications to the entanglement structure might reflect in the underlying spacetime geometry. In this flow prescription, the HEE is written as,
\begin{subequations}\label{def&constraints}
\begin{equation}\label{def}
	S_A = \max_{\upsilon} \int_A \upsilon ,
\end{equation}
  subject to the constraints:
  \beq\label{constraints}
   \nabla_{a} \upsilon^a = 0 \,; \quad |\upsilon| \le \dfrac{1}{4 G_N} .
  \eeq
\end{subequations}
   Here, $\upsilon^a$ is the bit thread vector field, which at the RT surface saturates the above inequality: $\upsilon^a |_{RT}= \mathrm{C} n^a$; $n^a$ being the unit normal of the RT surface and $\mathrm{C} =1/4 G_N$. The equivalence between this max-flow picture and the minimal-area RT formula is guaranteed by a continuous version of the max-flow min-cut theorem, where the RT surface $m(A)$ emerges as the bottleneck, or minimal cut, whose capacity dictates the maximum possible flux (see Appendix \ref{appsec:bithtreads} for a technical review).

 This elegant, flow-based perspective has turned out to be a useful approach for a variety of problems. It has been generalised to covariant settings, introduced by Headrick and Hubeny \cite{Headrick:2022nbe}, which accommodates the dynamics of entanglement in non-static spacetimes by employing a family of divergenceless flows consistent with the extremal surface prescription in the Hubeny-Rangamani-Takayanagi (HRT) framework \cite{Hubeny:2007xt} and beyond. Additionally, bit threads have been applied to study various information-theoretic quantities in holography. They have provided insights into the monogamy of mutual information \cite{Cui:2018dyq}, a fundamental constraint on the sharing of entanglement among multiple parties; as well as the entanglement of purification \cite{Du:2019emy, Bao:2019wcf, Harper:2019lff}, which quantifies correlations in mixed quantum states. The Riemannian bit threads have also been extended to cases like higher curvature gravity \cite{Harper:2018sdd}, quantum bit threads \cite{Agon:2021tia, Rolph:2021hgz}, PEE threads \cite{Lin:2023rxc}, and entanglement threads \cite{Lin:2025yko, Lin:2025jjh}, to name some of the few.
   
   The major thrust of this paper is to explore the relationship between bit threads and the rigorous framework of the covariant phase space (CPS) formalism \cite{Iyer:1994ys, Wald:1993nt}. Building on prior work that constructed perturbative bit threads using CPS methods \cite{Agon:2020mvu}, we investigate whether the HEE bit thread field can be directly identified with the conserved currents derived from CPS. Bit threads, constructed in this way, have been referred to as \textit{canonical bit threads}. Since the covariant phase space formalism is very general, bit threads written in this language offer a promising avenue for deeper insights into the geometric interpretation of concepts related to holographic entropy. However, there are certain challenges when we directly attempt to obtain the bit thread vector field from a current form derived using the CPS formalism. We find out that while CPS-inspired vector fields $v^a_{cps}$ are divergenceless, these generally do not satisfy the norm bound and saturation conditions required for bit threads. An extra term  $v_{\ed}^a$ (which does not change the cohomology class of the dual form) needs to be incorporated in many cases, except for certain situations like Rindler-AdS geometries, where the RT surface coincides with the horizon. But once we obtain the vector field in Rindler like coordinates, we can use coordinate transformations to find a valid thread configuration in the other coordinate system where we initially needed to add $v^a_\ed$ (like Poincar\'{e}) without any prior knowledge of the RT surface.
  
  This connection inspires us to explore whether the flow paradigm can account for other geometric entropy concepts. We argue that the Wald entropy (associated with black holes) and differential entropy (associated with spacetimes with a hole) can indeed be naturally expressed as the flux of specific CPS-derived flows linked to the relevant Killing vectors. Furthermore, the CPS method allows us to reformulate the first laws governing variations of these entropies in terms of associated thermodynamic flows, offering a unified geometric perspective for entropy.

  Finally, we address the incorporation of quantum effects, discussing the constraints it provides on bulk entanglement via energy conditions. We also outline the challenges and potential strategies for constructing covariant bit threads in certain dynamical spacetimes, which are helpful in studying processes such as thermalisation and black hole evaporation.
    
   The paper is organised as follows. We discuss the construction of a flow from the CPS formalism in section \ref{sec:btcps}. Then we apply the method to reinterpret the black hole entropy and differential entropy in terms of these threads in certain gravitational backgrounds in section \ref{sec:extension}. In section \ref{sec:first_laws}, we discuss the first laws of the entropies mentioned above in terms of bit threads. The quantum corrections and their relation to the dominant energy condition are addressed in section \ref{sec:quantum}.  In section \ref{sec:dynamical}, we discuss the bit thread construction in AdS-Vaidya spacetime. Finally, we conclude with the summary of the work and possible future directions in section \ref{sec:discussion}.

\section{Bit threads and CPS}\label{sec:btcps}
In general, it is difficult to obtain a bit thread configuration unless the given geometry is highly symmetric. There have been proposals for constructing these vector fields using geodesics \cite{Agon:2018lwq}; unit normals to the RT surfaces, a level-setting approach \cite{Agon:2020mvu}; or as magnetic flows using a modified Biot-Savart law \cite{Pinto:2022bho, Gursoy:2023tdx}. Usually, we need to know the metric to do calculations. This motivates us to seek a more fundamental approach, potentially one that is less reliant on the explicit metric and more connected to the geometry, intrinsic symmetries, and conserved structures of the gravitational theory. The CPS formalism, which operates naturally with differential forms and conserved currents, presents a promising avenue.\footnote{Another approach is to use calibrations, in which the RT surface is realised as a special Lagrangian cycle \cite{Bakhmatov:2017ihw}.} So, a natural question arises: \emph{Can we construct flows in a more canonical way using the CPS method?}\footnote{An attempt to construct a bit thread configuration using the Noether charge form was made in \cite{Oh:2017pkr}.} The authors of \cite{Agon:2020mvu} have already explored this direction for perturbative bit threads, terming them `canonical bit threads'. Here, we investigate how this idea can be extended to non-perturbative cases as well. The main trick is to use parametric variations instead of general perturbations within the CPS framework, which works equally well. Sometimes this is referred to as the `solution phase space' method \cite{Hajian:2015xlp}. For completeness, a short review of the CPS formalism is provided in appendix \ref{sec:cps}. Although there are various versions of the CPS formalism, we will mainly refer to that espoused by Iyer and Wald \cite{Iyer:1994ys}, as all of them lead to the same entropy formula, which is of our primary interest. First, we will sketch the general idea of how to obtain the bit thread vector field from the CPS, and then we will look into its applicability.

 In the CPS formalism, the spacetime is considered to be a $d$-dimensional manifold, and the phase space is a symplectic manifold. The Hamiltonian then is given by integrating the symplectic current $\bom$ (See \eqref{sympcurrent}-\eqref{hamiltonianIW}) over a Cauchy slice. This construction is not only restricted to the Hamiltonian, but is true for any conserved charge associated with a specific set of transformations generated by the symmetry generators as follows. The symplectic current is a closed $(d-1;2)$-form\footnote{In this notation, a $(p;q)$-form means it is a $p$-form in spacetime and $q$-form in phase space following \cite{Hajian:2015xlp}. We will frequently use the convention $p$-form, which is in the spacetime unless specified.}, i.e.,
\begin{equation}
	\ed \bom (\delta_1 \Phi, \delta_2 \Phi, \Phi) = 0,
\end{equation}
for any two perturbations satisfying the linearised equations of motion. Then, on the phase space $\cP$, one can locally define a $(d-2;1)$-form $\mathbf{k}_\zeta$ for any transformation generator $\zeta$ as
\begin{equation} \label{fundid}
	\bom (\delta \Phi, \delta_\zeta \Phi, \Phi) = \ed \mathbf{k}_\zeta  (\delta \Phi,  \Phi).
\end{equation}
The above equation is called the `\textit{fundamental identity}' of the CPS formalism \cite{Hollands:2012sf}. Variation of a conserved charge associated with $\zeta$ is given by,
\begin{equation}
	\delta H_\zeta (\Phi) = \int_\Sigma  \ed\mathbf{k}_\zeta (\delta \Phi, \Phi) = \int_{\partial \Sigma} \mathbf{k}_\zeta (\delta \Phi, \Phi).
\end{equation}
It is the same equation \eqref{hamiltonianIW} in terms of the $(d-2)$-form $\mathbf{k}$. 

\subsection{Codim-2 Current}\label{subsec:currentcps}

As mentioned earlier, there are different versions of the CPS formalisms, we will look into the codimension-2 current $\mathbf{k}_\zeta$,\footnote{Sometimes it is called the \emph{`surface charge'}, but we will stick to the terminology of codim-2 current as it is convenient to interpret when we construct a flow using it.} for two of them namely, the Iyer-Wald formalism and the covariant phase space with boundaries.

\subsubsection*{Codim-2 Current in Iyer-Wald CPS}

In the Iyer-Wald formalism \cite{Iyer:1994ys}, the symplectic current is given by,
\begin{equation}
	\bom_{\rm{IW}}(\delta_{1} \Phi, \delta_{2} \Phi,  \Phi) = \delta_{1} \mathbf{\Theta}_{\rm{IW}}(\delta_{2} \Phi, \Phi) - \delta_{2} \mathbf{\Theta}_{\rm{IW}} (\delta_{1} \Phi,  \Phi), \nonumber
\end{equation} 
and for a diffeomorphism invariant theory, it becomes
\begin{equation}
	\bom_{\rm{IW}}(\delta \Phi, \mathcal{L}_\xi \Phi,  \Phi) = \delta \mathbf{\Theta}_{\rm{IW}}(\mathcal{L}_\xi \Phi, \Phi) - \mathcal{L}_\xi \mathbf{\Theta}_{\rm{IW}} (\delta \Phi,  \Phi),
\end{equation}
with $\xi$ being the diffeomorphism generator and $\mathcal{L}_\xi$ is the Lie derivative w.r.t $\xi$. Here $\zeta$ is replaced with $\xi$ to differentiate between a diffeomorphism generator and any general transformation. To obtain an explicit expression for $\mathbf{k}$, the other ingredient is the Noether current, which is defined in terms of the diffeomorphism generator $\xi$,  symplectic potential $\mathbf{\Theta}_{\rm{IW}}$ and the Lagrangian $\boldsymbol{ \mathrm{L}}$ as given in equation \eqref{noethercurrent}: $\boldsymbol{\mathrm{J}}_{\xi} \equiv \mathbf{\Theta}_{\rm{IW}} (\mathcal{L}_\xi \Phi, \Phi) - \xi .\boldsymbol{ \mathrm{L}}$. Then the variation of the Noether current $\boldsymbol{\mathrm{J}}_{\xi}$ yields
\begin{equation}
	\delta \boldsymbol{\mathrm{J}}_\xi = \delta \mathbf{\Theta}_{\rm{IW}}(\mathcal{L}_\xi \Phi, \Phi) - \mathcal{L}_\xi \mathbf{\Theta}_{\rm{IW}} (\delta \Phi,  \Phi) + \ed \left( \xi.\mathbf{\Theta}_{\rm{IW}} (\delta \Phi,  \Phi) \right) \nonumber
\end{equation}
using the Cartan's magic formula: $\cL_V \boldsymbol{\alpha}= V.\ed\boldsymbol{\alpha} + \ed (V.\boldsymbol{\alpha})$ for any vector $V$ and differential-form $\boldsymbol{\alpha}$. Substituting the above expression of $\delta \boldsymbol{\mathrm{J}}_\xi$  in the definition of $\bom_{\rm{IW}}$, one gets
\begin{align}
	\bom_{\rm{IW}} &= \delta \boldsymbol{\mathrm{J}}_{\xi} - \ed (\xi.\mathbf{\Theta}_{\rm{IW}}) = \delta (\ed\boldsymbol{ \mathrm{Q}}_{\xi}) -  \ed (\xi.\mathbf{\Theta}_{\rm{IW}}), \nonumber \\
	&= \ed (\delta \boldsymbol{ \mathrm{Q}}_{\xi} - \xi.\mathbf{\Theta}_{\rm{IW}}).
\end{align}
Here, $\boldsymbol{ \mathrm{Q}}_{\xi}$ is the Noether charge defined in \eqref{noethercharge}. Now, comparing the above equation with the fundamental identity \eqref{fundid}, the codimension-2 current for diffeomorphism invariant theories is given by,
\begin{equation}
	\mathbf{k}_\xi = \delta \boldsymbol{ \mathrm{Q}}_{\xi} - \xi.\mathbf{\Theta}_{\rm{IW}} .
\end{equation}
If one can find a $(d-1)$-form $\bos{B}$ such that $\mathbf{\Theta}_{\rm{IW}}= \delta \bos{B}$ then the above equation can be rewritten as
\beq \label{kcur2}
\mathbf{k}_\xi=\delta \bos{\mathit{j}}_\xi,
\eeq
where the codim-2 form $\bos{\mathit{j}}_\xi$ is given by,
\beq
\bos{\mathit{j}}_\xi = \boldsymbol{ \mathrm{Q}}_{\xi} - \xi.\bos{B} \;.
\eeq
This current form is our main ingredient to construct a configuration for bit threads.

\subsubsection*{Codim-2 current in CPS with boundaries}

In the covariant phase space formalism with boundaries \cite{Harlow:2019yfa}, one starts with an action of the theory which has an extra boundary Lagrangian \eqref{actioncpsb} and follows a similar procedure as stated earlier. In this case, the (pre)-symplectic current is defined as,
\begin{equation}
	\bom_{\rm{HW}} = \delta \left(\mathbf{\Theta}_{\rm{HW}} - \ed \boldsymbol{ \mathsf{C}} \right)|_{\tilde{\mathcal{P}}}
\end{equation}
where, $\tilde{\mathcal{P}}$ denotes the pre-phase space. The symplectic current is closed: $\ed\bom_{\rm{HW}} = 0$. Pre-symplectic form is defined as the symplectic current integrated over a Cauchy surface, $\tilde{\mathbf{\Omega}}_{\rm{HW}} = \int_\Sigma \bom_{\rm{HW}}$. Then, the change in Hamiltonian is  given by
\begin{equation} \label{hamiltonian2}
	\delta H_\xi = - X_\xi . \tilde{\Omega}_{HW} = \int_\Sigma \left( - X_\xi . \bom_{\rm{HW}} \right), 
\end{equation}
where $\xi^\mu$ is the diffeomorphism generator and $X_\xi$ is a vector field defined on the configuration space given by, $X_\xi = \int d^{d}x \mathcal{L}_\xi \Phi^{i}(x) \frac{\delta}{\delta \Phi^i}$. In CPSB, the Noether current is expressed as
\begin{equation}
	\boldsymbol{\mathrm{J}}_{\xi} \equiv X_\xi.\mathbf{\Theta}_{\rm{HW}} - \xi.\boldsymbol{ \mathrm{L}}, \nonumber
\end{equation}
and the Noether charge is $\boldsymbol{ \mathrm{Q}}_\xi$ which satisfies $\boldsymbol{\mathrm{J}}_\xi = \ed \boldsymbol{ \mathrm{Q}}_\xi$ as $\ed \boldsymbol{\mathrm{J}}_\xi = 0$. Substituting them in the r.h.s of \eqref{hamiltonian2}, one gets
\begin{equation}
	- X_\xi . \bom_{\rm{HW}} = \delta \boldsymbol{\mathrm{J}}_\xi + \ed \left( \delta_\xi \boldsymbol{ \mathsf{C}} - \delta \left[ X_\xi . \boldsymbol{ \mathsf{C}} \right] - \xi.\mathbf{\Theta}_{\rm{HW}} \right). \nonumber
\end{equation}
Now, operating exterior derivative on the above quantity gives us
\begin{eqnarray}
	\ed\left( - X_\xi . \bom_{\rm{HW}} \right) = \delta \ed \boldsymbol{\mathrm{J}}_\xi + \ed^2 \left( \delta_\xi \boldsymbol{ \mathsf{C}} - \delta \left[ X_\xi . \boldsymbol{ \mathsf{C}} \right] - \xi.\mathbf{\Theta}_{\rm{HW}} \right) = 0, \nonumber
\end{eqnarray}
as $\boldsymbol{\mathrm{J}}_\xi$ is locally an exact form and $\ed^2 = 0$ by definition. It follows immediately that `$- X_\xi. \bom_{\rm}$' is a closed form. Locally, it can be made an exact form by Poincar\'{e}'s lemma, i.e., it can be expressed as an exterior derivative of some codim-2 form: $- X_\xi. \bom_{\rm{HW}} = \ed\tilde{\mathbf{k}}_\xi$, where $\tilde{\mathbf{k}}_\xi$ is given by,
\begin{equation}
	\tilde{\mathbf{k}}_\xi = \delta \boldsymbol{ \mathrm{Q}}_\xi + \delta_\xi \boldsymbol{ \mathsf{C}} - \delta \left[  X_\xi . \boldsymbol{ \mathsf{C}} \right] - \xi.\mathbf{\Theta}_{\rm{HW}} \, . \nonumber
\end{equation}
We can further simplify it by using the boundary conditions and Cartan's formula, which results in
\begin{equation}
	\tilde{\mathbf{k}}_\xi = \delta\left( \boldsymbol{ \mathrm{Q}}_\xi + \xi.\boldsymbol{{\ell}}  - X_\xi .\boldsymbol{ \mathsf{C}} \right).
\end{equation}
As seen from the above equation, we can find a current $\tilde{\bos{\mathit{j}}_\xi}$ such that $\tilde{\mathbf{k}}_\xi = \delta \tilde{\bos{\mathit{j}}_\xi}$, which is expressed as
\beq
\tilde{\bos{\mathit{j}}_\xi} =\left( \boldsymbol{ \mathrm{Q}}_\xi + \xi.\boldsymbol{{\ell}}  - X_\xi .\boldsymbol{ \mathsf{C}} \right) .
\eeq
This codimension-2 current can be used to construct various conserved charges. Now, we will examine how these current forms can be utilised to construct a flow for bit threads.

\subsection{Bit threads from CPS}\label{subsec:btcps}	
 In the previous subsection, we saw how one can find the $(d-2)$-form $\mathbf{k}_\xi$ (and $\bos{\mathit{j}}_\xi$) from the covariant phase space formalism. Here, we will show how to construct a flow from this form, which is a potential bit thread vector field. Both of the above-mentioned CPS formalisms are equally valid for this construction, as they lead to the same entropy. We will mainly follow the IW prescription. The codimension$-2$ current $\mathbf{k}_{\xi}$ is given by
\beq
	\mathbf{k}_\xi = \delta \bos{\mathrm{Q}}_\xi - \xi.\mathbf{\Theta}_{\rm{IW}} \,, \nonumber
\eeq
for any diffeomorphism generator $\xi^\m$. From hereafter, we will be omitting the suffix `IW' to simplify the notations. In the covariant phase space formalism \cite{Iyer:1994ys, Hajian:2015xlp}, these codim-2 currents are used to obtain variations of different conserved charges in a black hole background, which lead to the renowned first law of black hole thermodynamics (see appendix \ref{sec:bhthermodynamics} for a short review). We can extend this method to obtain the conserved charges by using the $(d-2)$-form $\bos{\mathit{j}}_\xi$ (if present) instead of the usual $\mathbf{k}_\xi$. We mainly focus on the exact symmetries (such as isometry) of the theory. When $\xi^\m$ is a Killing vector (or any exact symmetry generator), the symplectic current $\bom (\delta \Phi,  \mathcal{L}_\xi \Phi,  \Phi)$ vanishes, i.e.,
\beq
	\bom(\delta \Phi,  \mathcal{L}_\eta \Phi,  \Phi) = 0 = \ed \mathbf{k}_\eta (\delta\Phi, \Phi)= \ed \delta \bos{\mathit{j}}_\eta (\delta\Phi, \Phi), \nonumber
\eeq
where we have denoted `$\eta$' as the set of all Killing vectors. The above equation implies that $\mathbf{k}_\eta (\delta\Phi, \Phi)$ is a closed form and so is $\bos{\mathit{j}}_\eta$. Then in the presence of a Riemannian metric $g_{ab}$ ($a, b = 1,\cdots,d-1$), we can have a divergenceless vector field $v^a$ from either of the closed forms $\mathbf{k}_\eta (\delta\Phi, \Phi)$ or $\bos{\mathit{j}}_\eta $. We will use the latter one to obtain a vector field using the map:\footnote{ \textbf{On the uniqueness of $\bos{v_{cps}^{a}}$ :} The fundamental identity \eqref{fundid} allows us to define the codim-2 current $\mathbf{k}_{\xi}$ from the symplectic current, but this definition involves some ambiguities that we need to address. The symplectic potential $\mathbf{\Theta}$ itself is not uniquely defined \cite{Iyer:1994ys}. Adding boundary terms to the Lagrangian changes $\mathbf{\Theta}$ but leaves the symplectic current $\bom= \delta \mathbf{\Theta}$ unchanged. More generally, we can modify $\mathbf{\Theta} \rightarrow \mathbf{\Theta} - \mathrm{d}\mathbf{B}$, which transforms the symplectic current as $\bom \rightarrow \bom + \mathrm{d}\bom_{B}$. These are called ``\emph{residual ambiguities}'' and reflect our freedom in choosing boundary terms \cite{Compere:2018aar}. However, for exact symmetries (such as Killing symmetries), these ambiguities cancel out. When we transform the symplectic current $\bom \rightarrow \bom + \mathrm{d}\bom_{B}$, the codim-2 current transforms as:
	\beq
		\mathbf{k}_{\xi} \rightarrow \mathbf{k}_{\xi} + \bom_{B}(\delta \Phi, \mathcal{L}_\xi \Phi, \Phi) \nonumber
	\eeq
		For Killing vectors or generators of exact symmetries, the Lie derivative $\mathcal{L}_\xi \Phi$ equals the field variation $\delta_\xi \Phi$, so the additional term vanishes and $\mathbf{k}_{\xi}$ remains unchanged.
		This means the vector field $v_{cps}^a$ obtained through the CPS construction is uniquely defined (up to the usual degrees of freedom in $\mathbf{k}_\xi$) when we work with exact symmetries.\label{uniqueness}}
\beq \label{vj}
	v_{cps}^a = g^{ab} (\star \bos{\mathit{j}}_\eta )_b ,
\eeq
where, `$\star$' is the Hodge dual defined on a Cauchy surface.\footnote{We will mostly work in the spacetimes that are asymptotically AdS. AdS spacetime is not globally hyperbolic and therefore does not admit a Cauchy surface, which is required to have a well-defined initial value problem and hence the canonical phase space. However, a well-defined phase space can be constructed in AdS by choosing suitable boundary conditions \cite{Ishibashi:2004wx, Wald:1980jn}. So, we will use the term \emph{`Cauchy'} surface to refer to the spacelike hypersurfaces with proper boundary conditions in asymptotic AdS spacetimes.} Again, we can write \cite{Wald:1984rg},
\begin{equation}
	\ed \bos{\mathit{j}}_\eta  = (\nabla_{a} v_{cps}^a ) \boep,
\end{equation}
where $\boep$ is the volume form of the Cauchy slice. As $\bos{\mathit{j}}_\eta$ is a closed form, it implies that the vector field is indeed divergenceless (i.e., $\nabla_a v_{cps}^a = 0$) everywhere. Equivalently, we can have the relation at the boundary of a Cauchy surface as
\beq
\bos{\mathit{j}}_\eta \Big|_{\partial \Sigma} = (n_{a}v_{cps}^{a}) \tilde{\boep},
\eeq
where, $\tilde{\boep}$ is volume form on its boundary. We will use the $(d-2)$-form $\mathbf{k}$ to construct a vector field to study the thermodynamics of a gravitational system. We will discuss this briefly in the section \ref{sec:first_laws}.

   A central requirement of the bit thread formalism is that $v^a$ must be divergenceless. The vector fields derived from the CPS formalism using the codim-2 form $\mathbf{k}_\xi$ (or $\bos{\mathit{j}}_\xi$) automatically fulfil this condition, which suggests that imposing the remaining bit thread constraints on these CPS-derived flows might provide a canonical construction for bit threads. To have a better understanding of the connection between these two formalisms, we must re-examine the properties of the codim-2 form $\mathbf{k}_\xi$. Although $\mathbf{k}_\xi$ resolves ambiguities present in the symplectic current (as explained in the footnote \ref{uniqueness}), it can be expressed up to an exact form. This freedom in $\mathbf{k}_\xi$ can be utilised to satisfy the properties of the bit threads. So we can write the current form as
\beq
\mathbf{k}_{\xi}  \rightarrow \mathbf{k}_{\xi} + \ed \mathbf{k}_{B_\xi}  .\nonumber
\eeq 
This tells us that $\mathbf{k}_\xi$ belongs to the cohomology class $\left[\mathbf{k}\right] = \{\mathbf{k}_{\xi}\, | \, \ed \mathbf{k}_{\xi}= 0,\, \mathbf{k}_{\xi} \sim \mathbf{k}_{\xi} + \ed \mathbf{k}_{B_\xi} \}\in H_{dR}^{d-2}(\mathcal{M})$. Similarly, the current form $\bos{\mathit{j}}_\xi$ belongs to the cohomology class $\left[ \bos{\mathit{j}} \right]\in H_{dR}^{d-2}(\mathcal{M})$ given by,
\beq \label{freedom}
\bos{\mathit{j}}_\xi \sim \bos{\mathit{j}}_\xi' \quad s.t. \quad \bos{\mathit{j}}_\xi'-\bos{\mathit{j}}_\xi= \ed \bos{\mathit{j}}_{B_\xi} \,.
\eeq
The same ensues for the vector field obtained from the above current form: 
\beq
v^a = g^{ab} (\star \bos{\mathit{j}}_\xi)_b +  g^{ab} (\star \ed \bos{\mathit{j}}_{B_\xi})_b = v^a_{cps} + v_\ed^a,
\eeq
 where, $v^a_\ed$ is the vector field corresponding to the co-exact $1$-form $\star \ed \bos{\mathit{j}}_{B_\xi}$ and is divergence free. Now, if the given manifold (or the domain of interest) is simply connected, $v^a_\ed$ can be expressed as a gradient of some function $\varphi(x)$. So, the term $v_{\ed}^a = \nabla^a \varphi$ is like a $U(1)$ gauge freedom,\footnote{We thank Mohd Ali for emphasising the importance of this term and for valuable discussions.} which will help us to satisfy the boundary conditions for bit threads in some cases. We are mainly interested in the regions of the spacetime manifold where the domain is simply connected. Then, we can say that $v^a$ and $v^a_{cps}$ are from the same equivalence class:
\beq
 [v^a] = \Bigl\{ v^a_{cps}\, | \, v^a_{cps} \sim v^a_{cps} + \nabla^a \varphi \, \Bigl\}\, \quad \; ~\text{with} \; \nabla^2 \varphi = 0 \, .
 \eeq
  Evidently, $v^a$ is a divergenceless vector field, and we will show that this becomes a valid bit thread configuration once we impose the conditions \eqref{constraints}. If the manifold has non-trivial topologies (when the manifold is not simply connected), the bit threads field $v^a$ and the vector derived from CPS $(v^a_{cps})$ differ by a divergenceless vector field $v^a_\ed$. Therefore, in general the equivalence class of $v^a_{cps}$ can be written as: $[v^a_{gen}] = \{ v^a_{cps}\, | \, \, v^a_{cps} \sim v^a_{cps} + v^a_\ed \, \}$  with $ \nabla_a v^a_\ed = 0$.

 The bit thread vector field that we have obtained using the $(d-2)$-form $\bos{\mathit{j}}$ arising from the CPS formalism is called the `\emph{canonical bit threads}'. Although this formalism does not require any metric until we switch to the notion of a vector, the map \eqref{vj} precisely uses the metric of a Cauchy surface to construct the desired $v_{cps}^a$ from the differential form $\bos{\mathit{j}}$. And it is essential to invoke the non-trivial $v^a_\ed$ to satisfy all the conditions of bit threads. For example, we will look into AdS$_3$ spacetime in two different coordinate systems.
 
\subsubsection*{Poincar\'{e} AdS$_3$}\label{sec:pads_3}
 Here we will discuss the construction of bit threads in Poincar\'{e} $\rm{AdS}_3$ background for some boundary interval and show how naively applying the $v_{cps}^a$ fails to reproduce the HEE and we do need to add a non-trivial vector field $v_{\ed}^a$ to satisfy all the conditions of bit threads. The metric in Poincar\'{e} coordinates is given by,
 \beq\label{poincareads}
 ds^2 = \dfrac{L^2}{z^2} \left( -dt^2 + dz^2 +dx^2 \right),
 \eeq
 where, $L$ is the AdS radius, $x$ is the boundary coordinate and the boundary CFT is at $z=0$. We will use the bulk modular flow of the $t=0$ slice: $\xi^{bulk}_{modular}\vert_{t=0}=\left(\dfrac{\pi}{R}[R^2-z^2-x^2]\partial_{t},0,0\right)$, which is a Killing vector of the spacetime. To find $\mathbf{k}_\xi$, we use the expression \cite{Hajian:2015xlp}, 
 \begin{equation} \label{keh}
 k^{\mu\nu}_\xi = \dfrac{1}{16 \pi G_N}\left[ \left( \xi^{\nu} \nabla^{\mu} h -  \xi^{\nu} \nabla_{\sigma} h^{\mu\sigma} +  \xi_{\sigma} \nabla^{\nu} h^{\mu\sigma} + \dfrac{1}{2} h \nabla^{\nu} \xi^{\mu} - h^{\delta\nu} \nabla_{\delta} \xi^{\mu} \right) - (\mu \leftrightarrow \nu) \right] ,
 \end{equation} 
where, $\xi^{\mu}$ is a Killing vector, $ h^{\mu\nu} = g^{\mu\sigma} g^{\nu\gamma} \, \delta g_{\sigma\gamma}$ is the metric perturbation  and $h \equiv h^{\mu}_{\mu} = g_{\mu\nu} h^{\mu\nu}$ is its trace. These indices run from $0$ to $d-1$ for a $ d$-dimensional spacetime. The codimension-2 form is then given by,
\beq
 \mathbf{k}_\xi = \dfrac{\sqrt{-g}}{(d-2)! \,2!}\, \epsilon_{\mu\nu\sigma_{1}\cdots\sigma_{d-2}} \, k^{\mu\nu}_\xi dx^{\sigma_{1}} \wedge \cdots \wedge dx^{\sigma_{d-2}} . \nonumber
\eeq
To obtain the explicit expression of $ k^{\mu\nu}_\xi$, we need to choose some variation of the metric \eqref{poincareads}. There are different ways to calculate the variation of the metric $\delta g_{\mu\nu}$. One of the most useful and also convenient for our case is to use the parametric variation denoted as 
 \begin{equation}
 \hdel g_{\mu\nu} (P,Q) = \dfrac{\partial g_{\mu\nu}}{\partial P} \hdel P + \dfrac{\partial g_{\mu\nu}}{\partial Q} \hdel Q, \nonumber
 \end{equation}
 where $P$ and $Q$ are some parameters of the metric. For the present case \eqref{poincareads}, the only constant we have is the AdS radius $L$. So, the parametric variation looks like $ \hdel g_{\mu\nu} (L) = \dfrac{\partial g_{\mu\nu}}{\partial L} \hdel L$. Using this parametric variation, we find the only non-zero component of the form $\mathbf{k}_\xi$ :
 \beq
  k^{tz}_\xi=- \dfrac{\hdel L}{8 G_N} \dfrac{z}{L^3} \left( \dfrac{R^2-x^2}{R} \right) . \nonumber
 \eeq
 We can also find another codim-2 current $\bos{\mathit{j}}_{\xi}$ such that $\mathbf{k}_\xi=\hdel \bos{\mathit{j}}_\xi$ given by
 \beq
  \mathit{j}^{tz}_\xi = \dfrac{1}{16 G_N} \dfrac{z}{L^2}  \left( \dfrac{R^2-x^2}{R} \right) .
 \eeq
 Now using the relation \eqref{vj}, we obtain a divergenceless vector field $v_{cps}^a$ given as\footnote{We thank Manish Ramchander for pointing out a mistake in this expression at the earlier stage of the work.}
 \beq\label{v_cps-poincare}
  v_{cps}^a = \left(v_{cps}^z,v_{cps}^x\right) = \dfrac{\mathfrak{B}}{4 G_N}\; \dfrac{z^2}{2R} \big( R^2-x^2, 0 \big) ,
  \eeq
 where, $\mathfrak{B}$ is an overall constant and is a function of $L$. We have kept the factor $1/4G_N$ separate to keep track of the constraints \eqref{constraints}. From this expression, it is clear that $v_{cps}^a$ is divergenceless. But it fails to be proportional to the unit normal $n^a$ on the RT surface. Recall the RT surface in Poincar\'{e} coordinates is given by $x^2 + z^2 = R^2$, $R$ being the radius of the RT surface. Then the unit normal is $n^a = (n^z,n^x) = \frac{z_m}{R} (z_m, x_m)$, where $z_m$ and $x_m$ are the coordinates on the minimal surface. One can easily check that $v_{cps}^a \neq n^a/4G_N$ on the RT surface. So, $v^a_{cps}$ alone cannot reproduce the HEE. However, we can utilise the gauge freedom $v^a_\ed=\nabla^a \varphi$ associated with $v^a_{cps}$ to make it a bit thread vector field. To find $v_{\ed}^a$, we first solve $\nabla_a \nabla^a \varphi = 0$  in the Poincar\'{e} background, and then substitute $v^a_\ed$ with the gradient of $\varphi$. Solving Laplace's equation with proper boundary conditions gives us the harmonic function
  \beq
  \varphi =\dfrac{L^2}{4 G_N R}  \left(z - \dfrac{ \mathfrak{B} z^3}{6} +  \mathfrak{B} \dfrac{z x^2}{2} + \dfrac{x\sqrt{R^2-z^2}}{z^2} \left(z -  \mathfrak{B} z^3 \right) \right).
 \eeq
Taking the gradient of the above function and putting together with \eqref{v_cps-poincare}, we find the total vector field,
 \beq
 v^a = \dfrac{1}{4 G_N} \left( \dfrac{z^2}{R} +  \dfrac{ \mathfrak{B} z^2}{2R}(R^2-x^2-z^2) , \mathfrak{B} \dfrac{z^3 x}{R} + \dfrac{\sqrt{R^2-z^2}}{R} \left(z -  \mathfrak{B} z^3 \right) \right),
 \eeq
 which reduces to $n^a$ on the RT surface and also satisfies the property $|v|\le 1/4G_N$. So this is a flow which correctly reproduces the HEE of a single interval on the boundary of Poincar\'{e} AdS. To interpret the different parts of $v^a$, we can say that the $v_{cps}^a$ store all the information regarding the conserved charges of the theory as it is derived from the symplectic current, and $\nabla^a \varphi$ helps to satisfy the constraints of bit threads. From this example, it is clear that $v^a$ and $v^a_{cps}$ are from the same equivalence class, i.e., $v^a \sim v^a_{cps}$. 
 
 A similar treatment, which we used in the Poincar\'{e} background, can also be done for other backgrounds. As an example, we discuss the case of D$1$-brane in appendix \ref{sec:d1-brane}. There may be alternative methods to find $v_{\ed}^a$ in different spacetimes that are more convenient to use than the one presented here. In case of Poincar\'{e} $\rm{AdS}$, we can directly obtain a flow solely using the methods of characteristics starting from the divergenceless condition. We briefly discuss it in appendix \ref{sec:pdebt}. 
 
\subsubsection*{Rindler AdS$_3$}
 In the previous subsection, we have seen that it was required to use the freedom presented in the definition of $v_{cps}^a$ to satisfy all the conditions to be a flow even in a simple geometry like Poincar\'{e} AdS. So, a natural question arises: \textit{Can we find geometries where $v_{cps}^a$ is sufficient to be a flow without the aid of its gauge freedoms?} To answer this question, we need to go to the reference frame of an observer whose causal boundary is bounded by the RT surface. Then, in the natural coordinates of that observer, we can obtain a valid bit thread vector field from the CPS formalism. One can always map a spherical region of the boundary CFT to this kind of wedge in the hyperbolic foliation of the dual AdS spacetime \cite{Casini:2011kv}. For example, consider an accelerated observer in $\rm{AdS}_3$ spacetime.
  The reference frame of the observer is the Rindler-$\rm{AdS}_3$ geometry, 
  \beq\label{rindlerads}
   ds^2 = -\left( \dfrac{\rho^2}{L^2} -\mathtt{a}^2\right)d\tau^2 + \left( \dfrac{\rho^2}{L^2} -\mathtt{a}^2\right)^{-1}  d\rho^2  +\rho^2 du^2 \, ,  \qquad  u\in \mathbb{R}\, , ~ \tau \sim  \tau + 2\pi i,
  \eeq
 where $\mathtt{a}$ is the Rindler acceleration parameter, $\tau$ is the Rindler time and $\rho=\mathtt{a} L$ is the horizon. We use the parametric variation $ \hdel g_{\mu\nu} (\mathtt{a}) = \dfrac{\partial g_{\mu\nu}}{\partial \mathtt{a}} \hdel \mathtt{a}$ to obtain the only non-zero component of $k_\xi^{ab}$,
 \beq
  k_\xi^{\tau\rho} = \dfrac{1}{8\pi G_N}  \dfrac{\mathtt{a} \hdel \mathtt{a}}{\rho} . \nonumber
 \eeq
 Here, we have chosen the Killing vector to be $\xi^\m = (\partial_\tau, 0, 0)$. We can find a $\bos{\mathit{j}}$ such that $\mathbf{k}_\xi =  \hdel \bos{\mathit{j}}_\xi$, and its non-trivial component is given by,
   \beq
   j_\xi^{\tau\rho} = \dfrac{1}{16 \pi G_N }  \dfrac{\mathtt{a}^2}{\rho} .
  \eeq
  This form is defined in the whole spacetime. By projecting it on a Cauchy surface and using the map \eqref{vj}, we find
  \beq
   v^\rho  = \dfrac{1}{4G_N} \dfrac{ \mathtt{a}\, L}{ \rho} \sqrt{\dfrac{\rho^2 }{L^2}-\mathtt{a}^2} \,. \nonumber
  \eeq
  By definition, the thread lines start at the boundary subregion (here, the whole boundary) and end at the complement of that subregion and horizon (or any other codim-2 surface present in the spacetime). But the above vector $v^\rho$ is directed towards the boundary, as that is the direction in which $\rho$ increases. So the actual vector field will be
   \beq \label{flow_rindler}
    v^a = \left( -v^\rho , v^u\right) =- \dfrac{1}{4G_N} \left(\dfrac{ \mathtt{a}\, L}{ \rho} \sqrt{\dfrac{\rho^2 }{L^2}-\mathtt{a}^2}, 0\right).
   \eeq
  The above vector field is divergenceless by definition and satisfies the necessary conditions for a valid bit thread configuration: $|v| \leq 1/4G_N,$ and $v^a = -n^a/4G_N$ on the horizon\footnote{Here we have an extra minus sign because the unit normal is pointing outward at the horizon whereas the direction of the bit thread is towards the horizon.} of Rindler space (i.e., the RT surface). These conditions ensure consistency with the bit thread formalism (see equation \eqref{flow}), allowing us to conclude that this vector field constitutes a physically admissible bit thread flow.
  
\subsubsection*{Rindler AdS to Poincar\'{e} AdS}
 Here we will show how to obtain a bit thread vector field in Poincar\'{e} coordinates from the vector \eqref{flow_rindler} by using the coordinate transformations presented in appendix \ref{sec:coordinates}.  First, we must ensure that the metrics presented earlier and in the appendix are consistent to use the transformations. Comparing the metrics \eqref{rindlerads} and \eqref{Rindler}, we set the acceleration parameter $\mathtt{a}=1$ in \eqref{rindlerads} which yields
 \beq
 ds^2 = -\left( \dfrac{\rho^2}{L^2} -1\right)d\tau^2 + \left( \dfrac{\rho^2}{L^2} -1\right)^{-1}  d\rho^2  +\rho^2 du^2 , \nonumber
\eeq
  and the vector field \eqref{flow_rindler} becomes
    \beq \label{v_rindler}
  v^a = \left( -v^\rho , v^u\right) =- \dfrac{1}{4G_N} \left(\dfrac{ L}{ \rho} \sqrt{\dfrac{\rho^2 }{L^2}-1}, 0\right) = -\dfrac{1}{4G_N} \left(\dfrac{ L}{ \rho} \sqrt{\dfrac{\trho^2 }{L}}, 0\right) .
  \eeq
 We can write the above vector in different bases as
  \beq
   \vec{v} = v^a \partial_a = \underbrace{\left(-v^\rho \partial_\rho ,v^u \partial_u\right)}_{Rindler \,AdS} = \underbrace{\left(v^z \partial_z , v^x \partial_x \right)}_{Poincar\acute{e} \,AdS} \nonumber .
  \eeq
  From \eqref{v_rindler}, we see that $v^u =0$.  So, in Poincar\'{e} coordinates the vector will be
  \beq\label{v(z,x)}
   v^a = (v^z, v^x) = -v^\rho(z,x) \left(\dfrac{\partial z}{\partial \rho} ,  \dfrac{\partial x}{\partial \rho}\right).
  \eeq
  To find $v^\rho (z,x)$, we need to find $\trho$. Using the relations \eqref{rho,u-z,x} and \eqref{x,z}, we get
  \beq\label{trho}
  \trho = \dfrac{1}{2z} (L^2 -z^2 -x^2).
  \eeq
  Substituting it and the expression of $\rho = \rho(z,x)$ as \eqref{rho(x,z)} in \eqref{v_rindler} yields
  \beq\label{tv(z,x)}
   \bar{v}^\rho (z,x) = 4 G_N v^\rho (z,x) = \dfrac{\trho}{\rho} = \dfrac{(L^2 -z^2 -x^2)}{\sqrt{\left(L^2 +x ^2+z^2\right)^2 - 4L^2x^2}} .
  \eeq
  Now, using the coordinate transformations \eqref{x,z}, we find
 \begin{subequations}
\beq
	\dfrac{\partial z}{\partial \rho}  = - \dfrac{2z^2}{\sqrt{\left(L^2 +x ^2+z^2\right)^2 - 4L^2x^2}} \left( \dfrac{L^2 - x^2 +z^2}{L^2 -z^2 -x^2} \right) , 
\eeq
\beq
	\dfrac{\partial x}{\partial \rho} = - \dfrac{4z^2}{\sqrt{\left(L^2 +x ^2+z^2\right)^2 - 4L^2x^2}} \left( \dfrac{z\, x}{L^2 -z^2 -x^2} \right).
\eeq
\end{subequations}
Putting the above relations along with \eqref{tv(z,x)} in \eqref{v(z,x)} gives us a vector field in the Poincar\'{e} coordinates,
\beq
 v^a = \dfrac{2}{4 G_N}  \left(\dfrac{z}{\sqrt{\left(L^2 +x ^2+z^2\right)^2 - 4L^2x^2}}\right)^2 \left( L^2 - x^2 +z^2,  2 z\, x  \right). \nonumber
\eeq
After rearranging a little bit, we find
\beq\label{v_rind-poinc}
 v^a = \dfrac{1}{4 G_N L}  \left(\dfrac{2Lz}{\sqrt{\left(L^2 +x ^2+z^2\right)^2 - 4L^2x^2}}\right)^2 \left( \dfrac{L^2 - x^2 +z^2}{2L},  \dfrac{z\,x}{L}  \right) =\dfrac{1}{4 G_N L} v^a_{geodesic} \,,
\eeq
where, $v^a_{geodesic}$ is the symmetric flow constructed using geodesics in \cite{Agon:2018lwq}. This vector field is divergenceless, obeys the norm bound $|v|\leq\frac{1}{4G_N L}$, and at the RT surface reduces to $v^a|_{RT} = \frac{n^a}{4G_N L}$. Evidently, it is a valid thread configuration. It is surprising that only using CPS formalism and the coordinate transformations, we found a thread configuration in the Poincar\'{e} background without any prior knowledge of the RT surface, whereas in constructing the symmetric flow from geodesics, the knowledge of the RT surface was required as shown in \cite{Agon:2018lwq}.
  \begin{figure}[htbp]
	\centering
	\includegraphics[width=8cm, height=6cm]{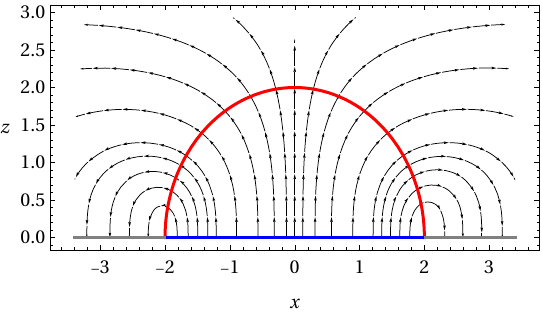}
	\caption{\textit{\label{fig:bt_rind-poin}Flow lines (black curves with arrows) for bit threads obtained from transforming a valid thread configuration in AdS-Rindler to Poincar\'{e}. The grey horizontal line denotes the boundary, and the red semicircle is the RT surface corresponding to the boundary subregion (shown in blue) $A\in [-2,2]$.}}
\end{figure}

So far, we have found a thread configuration for a fixed-length interval $L$, but in general, there may be any size of interval in the boundary CFT. In the bulk theory, we have corresponding RT surfaces of different radii. These RT surfaces are related via the isometries of the gravity theory. As discussed in appendix \ref{sec:coordinates}, one such simple isometry is boost \eqref{boost} in the embedding plane $(Y_{-1}, Y_2)$. We can find the vector field corresponding to one of these intervals by proceeding in a similar manner as above. In the boosted case, the expression \eqref{trho} becomes
\beq
  \trho = \dfrac{e^\beta}{2z} \left(e^{-2\beta}L^2 -z^2 -x^2\right).
\eeq
Using this and \eqref{rho(x,z)-boost} in \eqref{v_rindler} yields
\beq
  \bar{v}^\rho (z,x) = \dfrac{\trho}{\rho} = \dfrac{\left(e^{-2\beta}L^2 -z^2 -x^2\right)}{\sqrt{\left(e^{-2\beta}L^2 +x ^2+z^2\right)^2 - 4e^{-2\beta}L^2x^2}} .
\eeq
And incorporating the coordinate relations \eqref{x,z-boost}, we get
\begin{subequations}
\beq
	\dfrac{\partial z}{\partial \rho}  = - \dfrac{2e^{-\beta}z^2}{\sqrt{\left(e^{-2\beta}L^2 +x ^2+z^2\right)^2 - 4e^{-2\beta}L^2x^2}} \left( \dfrac{e^{-2\beta}L^2 - x^2 +z^2}{e^{-2\beta}L^2 -z^2 -x^2} \right) , 
\eeq
\beq
	\dfrac{\partial x}{\partial \rho} = - \dfrac{4e^{-\beta}z^2}{\sqrt{\left(e^{-2\beta}L^2 +x ^2+z^2\right)^2 - 4e^{-2\beta}L^2x^2}} \left( \dfrac{z\, x}{e^{-2\beta}L^2 -z^2 -x^2} \right).
\eeq
\end{subequations}
Combining all these in \eqref{v(z,x)}, we find 
\beq
 v^a =\dfrac{L}{4 G_N}  \left(\dfrac{2e^{-\beta}z}{\sqrt{\left(e^{-2\beta}L^2 +x ^2+z^2\right)^2 - 4e^{-2\beta}L^2x^2}}\right)^2 \left( \dfrac{e^{-2\beta}L^2 - x^2 +z^2}{2e^{-\beta}L},  \dfrac{z\,x}{e^{-\beta}L}  \right). \nonumber
\eeq
Now, if we identify $R=e^{-\beta}L$ to be the radius of the entangling surface in the boundary theory (or radius of the RT surface in the bulk), we end up with a vector field proportional to $v^a_{geodesic}$:
\beq
v^a = \dfrac{1}{4 G_N L}  \left(\dfrac{2 R z}{\sqrt{\left(R^2 +x^2+z^2\right)^2 - 4R^2x^2}}\right)^2 \left( \dfrac{R^2 - x^2 +z^2}{2R},  \dfrac{z\,x}{R}  \right).
\eeq
The flow lines of this vector field are shown in figure \ref{fig:bt_rind-poin}. The HEE corresponding to a boundary subregion $A$  in terms of the above thread configuration can be written as,
\beq
S_A = L \int_{m \sim A} v \,.
\eeq

\section{Extensions to other entropies}\label{sec:extension}
Bit threads can be related to other notions of entropy. In this section, we specifically attempt to relate it to black hole (BH) entropy and differential entropy. Since the CPS formalism is commonly applied in the context of black hole thermodynamics \cite{Wald:1993nt, Iyer:1994ys}, a natural extension is to derive a flow representation for black hole entropy; the same can be done for differential entropy \cite{Balasubramanian:2013rqa, Balasubramanian:2013lsa}, which is defined in both holographic and non-holographic spacetimes. 

\subsection{Black hole entropy}

  The entropy of a black hole has been a topic of interest for a long time, since Bekenstein \cite{Bekenstein:1973ur} to the present day. Among various approaches, the definition of black hole entropy as the Noether charge using the CPS formalism, developed by Wald \cite{Iyer:1994ys, Wald:1993nt}, is a well-established and standard notion of entropy, known as the `Wald entropy'. We will construct a flow corresponding to the Wald entropy of a black hole. Following the procedure described in section \ref{subsec:btcps}, we can find a divergenceless vector field. The only difference will be the norm bound at the RT surface. As we are interested in the black hole entropy, the norm bound should saturate on a minimal surface homologous to the black hole horizon (or, more specifically, the bifurcation surface). So the relevant condition becomes $v^a= - n_{\mathcal{H}}^a /4 G_N$ at the black hole horizon, $n_{\mathcal{H}}^a$ being the unit normal to the horizon. Similar to the AdS-Rindler case, the minus sign emphasises that the direction of $v^a$ is towards the horizon.\footnote{If we work with a Poincar\'{e} like coordinate system, then $v^a$ is naturally towards the horizon and we do not need to put the minus sign.} Then, the BH entropy is given by,
\begin{equation}
S_{BH}= \int_\cB v , \label{eq:bh_entropy}
\end{equation}
 where $v=\sqrt{\mathfrak{h}_\cB} |(n_{\mathcal{H}})_a v^a|$ as defined in appendix \ref{appsec:bithtreads}, $\cB$ is the bifurcation point of the black hole horizon, and $|v| \le 1/4 G_N$. Below, we will demonstrate how to find a flow for specific examples of black holes.
 
\subsubsection*{BTZ black hole}
   The metric of a BTZ black hole is given by, 
     \begin{equation}
     	ds_{\rm{BTZ}}^2 = - \mathcal{N}^2 dt^2 + \mathcal{N}^{-2} dr^2 + r^2 \left( \mathcal{N}^{\phi} dt + d\phi \right)^2  \nonumber
     \end{equation}
 where, $\mathcal{N}$ denotes the `lapse' function and $\mathcal{N}^{\phi}$ is the angular shift given by,
   \begin{eqnarray}
   	\mathcal{N}^2 = -M + \dfrac{r^2}{l^2} +\dfrac{J^2}{4 r^2} \; ; \quad
   	\mathcal{N}^{\phi} = - \dfrac{J}{2 r}. \nonumber
    \end{eqnarray}
 Here, `$M$' and `$J$' are the mass and angular momentum of the BTZ black hole, respectively; $l$ is the AdS radius. For simplicity, we will consider a non-rotating BTZ black hole, i.e.,  $J= 0$ and $\mathcal{N}^{\phi} =0$.  Now, the metric becomes,
  \begin{equation}
  	ds_{\rm{BTZ}}^2 = - \mathcal{N}^2 dt^2 + \mathcal{N}^{-2} dr^2 + r^2 d\phi^2 
  \end{equation}
 with $\mathcal{N}^2 = -M + \dfrac{r^2}{l^2}$. The horizon of this black hole is at $r_{\mathcal{H}} = l \sqrt{M}$. Consider the metric to be the function of mass of the BTZ black hole, i.e., $g_{ab} = g_{ab} (M)$. Then the parametric variation becomes,
 \begin{equation}
 \hdel g_{ab} (M) = \dfrac{\partial g_{ab}}{\partial M} \hdel M. \nonumber
 \end{equation}
 The different components of this variation are given by,
 \begin{equation}
 \hdel g_{tt} = \hdel M \;\; ; \;\; \hdel g_{rr} = \mathcal{N}^{-4} \hdel M = \left(  \dfrac{r^2}{l^2} - M \right)^{-2} \hdel M; \;\; \hdel g_{\phi\phi} = 0. \nonumber
 \end{equation}
 So, the only non-zero components of $h^{ab}$ are,
 \begin{equation}
 h^{tt} = g^{tt} g^{tt} \hdel g_{tt} = \mathcal{N}^{-4} \hdel M = \left(  \dfrac{r^2}{l^2} - M \right)^{-2} \hdel M; \nonumber
 \end{equation}
 and,
 \begin{equation}
 h^{rr} = g^{rr} g^{rr} \hdel g_{rr} = \mathcal{N}^{4} \mathcal{N}^{-4}  \hdel M  = \hdel M. \nonumber
 \end{equation}

 Evidently, $h^{ab}$ is traceless, $h=h^{a}_{a}=0$. Now, the antisymmetric nature of $k^{ab}$ tells us that all the diagonal components are zero, i.e., $k^{tt} = k^{rr} = k^{\phi\phi} = 0$. The only probable non-zero components are $k^{tr}$, $k^{t\phi}$, $k^{\phi r}$, and their antisymmetric pairs. Using \eqref{keh} it is straightforward to get the non-zero component $k^{tr} = \dfrac{1}{16\pi G_N r} \hdel M$, for the Killing vector $\xi^a = (\partial_t, 0, 0)$. For this parametric variation, we can easily find an expression for $\bos{\mathit{j}}_\xi$ given by,
  \begin{equation}\label{j-btz}
 	 j_\xi^{tr} = \dfrac{1}{16\pi G_N r} M.
 \end{equation}
 Now using the mapping \eqref{vj} and the boundary conditions $\big($i.e., $|v|\le \frac{1}{4 G_N}$ and $v^a=-\frac{1}{4 G_N} n_{\mathcal{H}}^a$ at the black hole horizon$\big)$, we find a vector field,
 \beq\label{v_cps-btz}
  v^a = \left(- v^r,v^\phi \right) =- \dfrac{1}{4 G_N}\left( \dfrac{\mathcal{N} M}{\kappa \, r}, 0 \right),
 \eeq
 where, $\kappa =r_{\mathcal{H}}/l^2$ is the surface gravity of the horizon. This vector field correctly reproduces the entropy of the BTZ black hole, indicating that it is a flow corresponding to a valid bit thread configuration for the black hole (or Wald) entropy. Generalisation to the rotating BTZ black hole is straightforward.

\subsubsection*{Schwarzschild-AdS black hole}
A similar procedure can be done for other black holes. Here, we consider a $(d+2)$-dimensional Schwarzschild-AdS black hole. The metric is given by,
\begin{align} \label{sch}
  ds^2_{\rm{Sch}} & = - f(r) dt^2 + \dfrac{dr^2}{f(r)} + r^2 d\Omega^2_{d} \nonumber \\
  & = - f(r) dt^2 + \dfrac{dr^2}{f(r)} + r^2 d\phi^2 + r^2 d\Omega^2_{d-1}, 
\end{align}
where, $f(r) = 1 - \dfrac{2 M}{r} + r^2$ with AdS radius $l= 1$. Considering spherical symmetries, we can suppress the last term in \eqref{sch}. Now the metric becomes,
\begin{equation} \label{esch}
	 ds^2_{\rm{Sch}}  = - f(r) dt^2 + \dfrac{dr^2}{f(r)} + r^2 d\phi^2 .
\end{equation}
The rest of the procedure is similar to the last case. Like the previous one, the Schwarzschild-AdS solution is also a solution of Einstein-Hilbert gravity. Then we can use \eqref{keh} to have the codim-2 current. Parametric variation gives us the different components of the variations of the metric as follows,
 \beq
  \hdel g_{tt} = \dfrac{2}{r} \hdel M ~ ; ~ \hdel g_{rr} = \dfrac{2}{r f(r)^2} \hdel M ~; ~ \hdel g_{\phi \phi} = 0. \nonumber
 \eeq
These lead us to the metric perturbations,
\beq
 h^{tt} =  g^{tt} g^{tt} \hdel g_{tt} =  \dfrac{2}{r f(r)^2} \hdel M , \nonumber
 \qquad
  h^{rr} =  g^{rr} g^{rr} \hdel g_{rr} = \dfrac{2}{r} \hdel M. \nonumber
\eeq
In this case, $h^{ab}$ is traceless: $h^a_a = h = 0$. Now returning to $\mathbf{k}_\xi$, the antisymmetric nature tells us that $k^{tt} = k^{rr} = k^{\phi\phi} = 0$. Moreover, we find that 
$k^{r\phi} = 0 = k^{t\phi}$. The only non-zero component is $k^{tr} = \dfrac{1}{8 \pi G_N r^2}  \hdel M$. And the $\bos{\mathit{j}}_\xi$ is given by,
  \begin{equation}
 	 j_\xi^{tr} = \dfrac{1}{8\pi G_N r^2} M.
 \end{equation}
 This $d$-form or the corresponding vector $v^a$ related to it via \eqref{vj} gives the black hole entropy upon integration on a codimension-2 surface homologous to the horizon (more specifically, the bifurcation point).
 
  We can think of these flows as the conserved current corresponding to different gravitational charges, and the divergenceless condition is the condition for conservation of those charges, similar to the continuity equation.

\subsection{Differential entropy}
 In the AdS/CFT correspondence, entanglement entropy in the boundary theory is computed via the Ryu-Takayanagi prescription, which asserts that the entanglement entropy of a boundary region is equal to the minimum area of a surface in the bulk spacetime. This gives us important insights into how quantum information is encoded in spacetime itself. The entanglement entropy extends naturally from individual intervals to overlapping boundary intervals. Each of them has an entanglement entropy given by the Ryu-Takayanagi formula. Taking the continuum limit of differences between these entropies, one introduces differential entropy \cite{Balasubramanian:2013lsa, Headrick:2014eia}:
\beq
S_{diff} = \int d\lambda \frac{\partial S\big(\gamma_L(\lambda),\gamma_R(\lambda)\big)}{\partial \lambda},
\eeq
where $\gamma_{L,R}(\lambda)$ are the common endpoints of a family of intervals parameterised by $\lambda$. For particular circumstances, the above integral provides the gravitational entropy of a bulk ``hole". Namely, differential entropy happens to be equal to the area of the hole divided by $4G_N$, 
\beq\label{diffentropy}
S_{diff} = \frac{\cA (\rm hole)}{4G_N}.
\eeq
This discovery indicates a profound relationship between measures of boundary entanglement and bulk geometry. The differential entropy agrees with the Bekenstein-Hawking entropy, which is an implication of the area law, strengthens the holographic principle and demonstrates that the intricate entanglement pattern on the boundary contains the same information as the bulk geometry. Moreover, differential entropy is a versatile tool for the investigation of bulk causal wedges and the reconstruction of bulk regions from boundary data. It provides insightful perspectives on how entanglement warps spacetime and enhances our knowledge of gravitational thermodynamics and holography. 

 Here, we show that the differential entropy corresponding to a hole in a spacetime can be obtained using the flow prescription in spherical Rindler and spherical Rindler-AdS spacetimes. The vector field is derived from CPS formalism with constraints $|v| \le 1/4 G_N$, $|v^a n_a |= 1$ at the boundary of the hole, and is divergenceless by definition. So, we can rewrite the differential entropy in terms of the threads as
\beq
S_{diff} = \int_{\mathfrak{b}} v,
\eeq
where $\mathfrak{b}$ is the boundary of the hole and $v = \sqrt{\mathfrak{h_b}}\,|n_a v^a|$. Now, we will look into the explicit construction of the vector field in spherical Rindler space.
 
\subsubsection*{Spherical Rindler space}
The concept of spherical Rindler space generalises the foundational framework of conventional Rindler space, which describes a family of observers undergoing radial acceleration away from a shared central point in Minkowski spacetime, establishing causal isolation from a spherical region of radius $R_0$ \cite{Balasubramanian:2013rqa}. Unlike linear Rindler acceleration, which produces planar horizons, this spherical generalisation creates an observationally inaccessible domain bounded by a perfectly symmetrical horizon surface.
The spherical Rindler horizon manifests as a topological boundary that separates the causally connected exterior region from the interior zone, which remains perpetually hidden to all accelerated observers. This horizon's geometrical properties maintain exact spherical symmetry, with its characteristic scale precisely determined by the parameter $R_0$, which defines the radius of the inaccessible central region. The $(d+1)$-dimensional metric is given by \cite{Balasubramanian:2013rqa}:
\beq
ds^2 = -r^2 dt^2 + dr^2 + (R_0 +r \cosh{t})^2 d \Omega_{d-1}^2 \, ,
\eeq
where, $d \Omega_{d-1}^2$ denotes the metric on a unit $(d-1)$-sphere. This metric structure reveals the intrinsic relationship between the acceleration parameter $R_0$, the radial coordinate $r$, and the temporal component that governs the proper time experience of the spherical Rindler observers. The coordinate singularity at $r=R_0$ mathematically encodes the causal boundary beyond which no information can propagate to the accelerating observers. The authors of \cite{Balasubramanian:2013rqa} showed that this horizon gives rise to a notion of entropy proportional to the area of the horizon, which was later named as `differential entropy'. To obtain the entropy associated with this hole, we generally take the near-horizon limit $r \rightarrow 0$. The metric approaches in the limit
\beq\label{NHsphrindler}
ds^2 = -r^2 dt^2 + dr^2 + R_{0}^2 \,d \Omega_{d-1}^2 \,.
\eeq
Now we can find a constant time slice in this geometry and employ the Iyer-Wald formalism to find the entropy of the hole, which is essentially given by \eqref{diffentropy}. We can also find a flow or bit thread configuration by following the procedure discussed in section \ref{subsec:btcps}. In the near-horizon geometry, we take the radius of the hole $R_0$ as the parameter to be used for the parametric variation of the metric. Using \eqref{keh} and \eqref{kcur2}, we find ,
\beq
k^{tr}_{\xi} = \dfrac{\hdel \mathfrak{c}}{16 \pi G_N r}\; ; \quad j^{tr}_{\xi} = \dfrac{\mathfrak{c}}{8 \pi G_N r},
\eeq
for the near-horizon geometry of the spherical Rindler space. Here $\mathfrak{c}$ is a function of $R_0$. Now using the relations \eqref{kcur2} and \eqref{vj}, we find the vector field,
\beq\label{flowhole}
 v^a = \left( -v^r,v^\Omega \right) = -\dfrac{1}{4 G_N} \left( \dfrac{\mathfrak{c}}{\kappa}, 0 \right) \, .
\eeq
 Notice that the vector field is constant along the $r$ direction, and the minus sign tells that it ends on the horizon of the hole. This reproduces the entropy of the hole as given by \eqref{diffentropy}. Therefore, we can see that it is possible to construct a flow-based picture for differential entropy. Moreover, from the above construction, we can say that the CPS formalism can also be applied to obtain the entropy of a hole in a spacetime. Both of these methods are applicable in the near-horizon limit, and the interpretation of the bit thread vector field away from the horizon remains an open question.

\subsubsection*{Spherical AdS-Rindler space}
 Spherical AdS-Rindler space can be viewed as a spherical Rindler space embedded in an asymptotically AdS spacetime. This space is holographically dual to the ultraviolet part of the boundary field theory, described by a finite time strip \cite{Balasubramanian:2013lsa}. This enables us to study a hole in a spacetime holographically. Analogous to the spherical Rindler space, this spacetime also has a family of accelerated observers whose worldlines are causally disconnected from a spherical hole in AdS. The metric of such spacetime is given by \cite{Balasubramanian:2013lsa},
 \beq
 ds^2 = l^2 \left( - \sinh^2{r} dt^2 +dr^2 \right) +\left( R_0 \cosh{r} + \sqrt{R_0^2 +l^2} \sinh{r}\cosh{t} \right)^2 d \Omega_{d-1}^2 \, ,
 \eeq
where $R_0$ is the coordinate radius of the hole. For convenience, we can take the AdS radius $l$ to be $1$. The horizon is at $r=0$, similar to the spherical Rindler case. Taking the near horizon limit $r\rightarrow 0$, we end up with same metric \eqref{NHsphrindler}:
\begin{equation*}
ds^2 = -r^2 dt^2 + dr^2 + R_{0}^2 \,d \Omega_{d-1}^2 
\end{equation*}
in the leading order. So, in this geometry, the flow will be the same as given in \eqref{flowhole}. It reproduces the gravitational entropy of the hole, viz., `differential' entropy. This method is only applicable in the near-horizon limit of the hole. If one wants to construct a thread configuration beyond the near-horizon limit, one needs to use the notion of covariant bit threads as formulated in \cite{Headrick:2022nbe}. 

\section{First laws of entropy using bit-threads}\label{sec:first_laws}

The CPS formalism helps us to construct a representative flow for a valid bit thread configuration. One can wonder if it is possible to represent different thermodynamical laws in the thread picture as well. In this section, we will mainly focus on the first law of thermodynamics for the entropies mentioned above. Before proceeding further, recall that the symplectic current vanishes for Killing symmetries, i.e., $\bom = 0 = \mathrm{d} \mathbf{k}_\eta$. So, we can construct a vector field from the codim-2 current as
\begin{equation} \label{vk}
	v^a_{th} = g^{ab} (\star \mathbf{k}_\eta )_b \, ,
\end{equation}
which is similar to the map \eqref{vj}, and ensures that $v^a_{th}$ is divergenceless. We can also impose a norm bound as we did for $v^a$ in the previous section. Then we end up with a `\emph{thermodynamic variation flow}'. This flow, when integrated at the boundary of a Cauchy surface, results in variations of different thermodynamic quantities.

\subsection*{First law of BH thermodynamics}

For a stationary, rotating, charge-neutral, axially symmetric black hole with non-degenerate Killing vectors and a bifurcation surface, we can derive the first law of black hole thermodynamics using the covariant phase space formalism as briefly discussed in appendix \ref{sec:bhthermodynamics}. For such black holes, the Killing vectors associated with the horizon, time translation and rotation are related via the equation \eqref{killingrelation},
\begin{equation}
	\xi_H = \xi_M - \Omega_{H} \xi_{J}, \nonumber
\end{equation}
where, $\Omega_{H}$ represents the angular velocity of the horizon. Now, it is easy to find out the relation between the codim-2 currents corresponding to the Killing vectors, and it is given by,
\beq\label{krelation}
\mathbf{k}_{\rm{\eta_S}} = \frac{2 \pi}{\kappa} \left( \mathbf{k}_{\rm{\eta_M}}  - \Omega_{\rm{H}} \mathbf{k}_{\rm{\eta_{J}}} \right),
\eeq
where we have used the definitions \eqref{Killingredefn} and relation \eqref{killingrelation2} with $\kappa$ being the surface gravity of the horizon. For a stationary, axisymmetric black hole, these flows correspond to different conserved charges similar to the definitions \eqref{canonicaldefn}:
\begin{subequations}
\begin{equation} 
	\delta M := \oint_{\mathcal{S}_\infty} v_{th,M} \;  ; \quad \delta J := \oint_{\mathcal{S}_\infty} v_{th,J} \; ; 
\end{equation}
\beq
	\delta S := \oint_\mathcal{B}  v_{th,S}\,.
\eeq
\end{subequations}
Here we have used the notation \eqref{flux}. For simplicity, we will omit the suffix `$th$' hereafter. These fields also follow a similar relation like \eqref{krelation}:
\beq \label{vrelation}
v^{a}_{S} = \frac{2 \pi}{\kappa} \left( v^{a}_{M} - \Omega_{H} v^{a}_{J} \right).
\eeq
 Looking at the above equation carefully, it is evident that only one vector field is sufficient to construct the different conserved charges of a black hole. Now we will proceed the same way as we did while deriving the first law of BH thermodynamics. We start with the fact that for Killing vectors, the symplectic form vanishes,
\beq
	\int_{\Sigma} \bom = 0 = \oint_{\partial \Sigma} \mathbf{k}_{\rm{\eta_S}} \, .
\eeq
 Now to construct a perturbative vector field, we can use \eqref{vk}. Then the above equation becomes,
 \beq
 \oint_{\partial \Sigma} \big( n_{a} v^{a}_{S}\big) \bos{\tilde{\epsilon}} = 0 \qquad
	\Rightarrow \qquad \int_{\Sigma} \big( \nabla_{a} v^{a}_{S}\big) \bos{\epsilon} = 0.
\eeq

Here, $\bos{\tilde{\epsilon}}$ denotes the volume form of the boundary of the Cauchy surface $\partial \Sigma$ and $\bos{\epsilon}$ denotes the volume form of the Cauchy slice $\Sigma$. As the volume form $\bos{\epsilon}$ can vary according to the size of the black hole and other geometrical properties, the integrand should vanish. This gives us a `local' 1st law of black hole thermodynamics as,
\beq
\nabla_{a} v^{a}_{S} = 0.
\eeq
So, the flow $v^{a}_{S}$ acts as a conserved current in the thermodynamics of black holes. The most direct interpretation of this local law is that certain concepts related to the thermodynamic quantities associated with the black hole are locally conserved. It implies that changes in these quantities are not merely due to fluxes across the horizon but are governed by a local balance equation within the bulk itself. This provides a more granular understanding of how black hole thermodynamics works.

\subsection*{First law of entanglement entropy}
Here, we will discuss how we can rewrite the first law of entanglement entropy in terms of bit threads. Similar to the previous section, we will use the canonical bit threads to construct a perturbative thread configuration in holographic spacetime. This approach was applied in \cite{Agon:2020mvu} to reconstruct the metric and to show how linearised Einstein equations can be obtained from perturbative bit threads. 

The first law of entanglement entropy says that the change in entanglement entropy is equal to the change in modular energy in the first order. In \cite{Faulkner:2013ica}, authors showed how one can extend the Iyer-Wald formalism in the subregion duality to reformulate the first law of entanglement entropy in a local way. The change in entanglement entropy and the modular energy can be written in terms of the codim-2 form $\mathbf{k}$ as follows \cite{Faulkner:2013ica}:
\beq
 \delta S_{A} = \int_{m(A)} \mathbf{k}_\xi\; ; \quad \delta E_{A} = \int_{A}  \mathbf{k}_\xi,
\eeq
where, subscript $A$ denotes the boundary subregion for which we are looking for the EE, $m(A)$ is the minimal surface homologous to $A$ and $\delta E_A$ denotes the change in modular energy. The on-shell condition  $\bom = 0 = \mathrm{d} \mathbf{k}_\xi$ leads to the first law $\delta S_{A} = \delta E_{A}$.\footnote{It can be easily checked as follows
\beq
  \int_{\Sigma_A}  \mathrm{d} \mathbf{k}_\xi = 0 = \int_{\partial\Sigma_A} \mathbf{k} _\xi \qquad
  \Rightarrow   \int_{m(A)} \mathbf{k}_\xi = \int_{A}  \mathbf{k}_\xi \qquad
	\Rightarrow \; \delta S_{A} = \delta E_{A} \nonumber \,.
\eeq} Now using the relation \eqref{vk}, we can rewrite the first law of entanglement entropy in terms of the bit thread vector fields,
\beq
\int_{\Sigma_A} \big( \nabla_{a} v^{a}_{th}\big) \bos{\epsilon_{\Sigma_A}} = 0 \, ,
\eeq
where $\Sigma_A$ denotes the region in between $A$ and $m(A)$, and $\bos{\epsilon_{\Sigma_A}} $ is the volume form of that region. Now, by construction, this perturbative bit thread is divergenceless, i.e., the integrand in the above equation does not depend on the size of the region $\Sigma_A$. So the first law of the entanglement entropy can be expressed in terms of the bit threads locally as:
\beq
 \nabla_{a} v^{a}_{th} = 0.
\eeq
Here, thermodynamic flow acts like a conserved current, ensuring that the EE and the modular energy are locally conserved.

\subsection*{First law of differential entropy}  
 In a similar way, we can extend the above procedure for differential entropy. But there are a few subtleties that need to be considered. Unlike the previous cases, the symplectic current does not vanish and gives rise to a volume term in the first law of causal diamond \cite{Jacobson:2018ahi}. Using the $``Complexity\,=\, Volume"$ conjecture, the authors of \cite{Sarkar:2020yjs} showed that one can obtain a first law of differential entropy in the holographic scenarios. So the perturbative flow corresponding to the differential entropy will no longer be divergenceless.  Unlike the EE and Wald entropy cases, the diffeomorphism generator here is a conformal Killing vector, which leads to the non-vanishing of the symplectic current. So, the fundamental identity \eqref{fundid} becomes,
 \beq
	\bom (\delta \Phi, \delta_\chi \Phi, \Phi) = \ed \mathbf{k}_\chi  (\delta \Phi,  \Phi),
 \eeq
 where $\chi$ is the conformal Killing vector. The l.h.s. gives rise to the variation of volume of the spatial sphere (hole) whose causal diamond is under consideration, and the r.h.s. gives rise to the variation of the area of the boundary of that spatial sphere \cite{Jacobson:2018ahi}.  Integrating over that spatial spherical region $\sigma$,
 \beq
  \int_{\sigma}	\bom (\delta \Phi, \delta_\chi \Phi, \Phi) = \int_\sigma \ed \mathbf{k}_\chi  (\delta \Phi,  \Phi) = \int_{\partial\sigma}  \mathbf{k}_\chi  (\delta \Phi,  \Phi).
 \eeq
  We can obtain a thermodynamic flow from $\mathbf{k}_\chi$, which is a perturbative Riemannian thread. On the other hand, we can construct a perturbative Lorentzian thread configuration from the symplectic current using the map: $\delta \mathfrak{u}^\m = g^{\m\n}(\ast \bom)_n$ as discussed in \cite{Pedraza:2021fgp}, where $\ast$ denotes the Hodge dual in a Lorentzian manifold. In terms of the threads, the above equation becomes,
\beq
 \int_{\sigma}	\delta \mathfrak{u} = \int_{\partial\sigma} v_{th} = \int_{\sigma}	\nabla_{a} v^{a}_{th}.
\eeq
This gives rise to the first law of differential entropy (or causal diamond). Locally, we can write
 \beq
 \delta \mathfrak{u} = \nabla_{a} v^{a}_{th} .
 \eeq
 The above equation provides a relation between the Lorentzian and Riemannian (perturbative) bit threads. 
 
\section{Quantum corrections}\label{sec:quantum}
We have mainly discussed the canonical bit threads in classical settings. In this section, we will include the quantum corrections, then use the positivity of the relative entropy to obtain an energy condition that imposes constraints on the quantum part. First, we will briefly review the relative entropy and the quantum bit threads.

\subsubsection*{Relative entropy}
In gravitational settings, relative entropy is defined as \cite{Lashkari:2015hha},
\beq
S_{\rm{rel}} \left(\rho_{A}^{\lambda} \parallel \rho_{A}^{0} \right) =  S_{\rm{rel}} \left(g^{\lambda} \parallel g^{0} \right) =  H_{\rm{bulk}}(g^{\lambda}) - S_{\rm{HEE}}(g^{\lambda}).
\eeq
where, $ H_{\rm{bulk}}$ is the bulk modular Hamiltonian or the gravitational energy,\footnote{It is same as the modular energy $E$ discussed in the previous section.} $\lambda$ denotes a one-parameter family. Then the positivity of relative entropy tells us that
\beq \label{positive}
\frac{d  S_{\rm{rel}}}{d\lambda} \Big{|}_{\lambda = 0} \, \ge 0\, , \nonumber
\eeq
which, in gravitational terms, is given by
\beq
\dfrac{d}{d\lambda} \left[H_{\rm{bulk}}(g^{\lambda}) - S_{\rm{HEE}}(g^{\lambda})\right]  \ge 0 \,.
\eeq
Again, by using the CPS formalism, one can rewrite the above inequality consisting of the gravitational energy and entropy as \cite{Lashkari:2015hha},
\beq
\dfrac{d}{d\lambda} H_{\rm{bulk}}(g^{\lambda}) = \int_A \mathbf{k}_\xi ~; \qquad \dfrac{d}{d\lambda} S_{\rm{HEE}}(g^{\lambda}) = \int_{m(A)} \mathbf{k}_\xi\,.
\eeq
Therefore, the positivity of relative entropy \eqref{positive} can be expressed in terms of $\mathbf{k}$ as follows
\begin{align}\label{positivity}
	&\int_A \mathbf{k}_\xi - \int_{m(A)} \mathbf{k}_\xi \; \ge 0 \nonumber \\
	\Rightarrow ~ & \int_{\Sigma_A} \mathrm{d} \mathbf{k}_\xi \; \ge 0.
\end{align} 
The boundary of the region $\Sigma_A$ is chosen to be: $\partial \Sigma_A =  A-m(A) $. The above relation will provide us with an energy condition, along with incorporating the concept of quantum bit threads.

\subsubsection*{Quantum bit threads} 
Quantum bit threads have been introduced recently in \cite{Agon:2021tia}, and independently in \cite{Rolph:2021hgz}. Although both are closely related and can be shown to be the same in leading order, here we will briefly discuss the former one. The quantum bit thread vector field $v$ is defined as $v = v_{0} + v_{q}$, such that:
\begin{subequations}
\beq
\int_{A} v = \int_{A} (v_0 + v_q) = S_{\rm{HEE}}^0 + S_{\rm{bulk}} ,
\eeq
but,
\beq
\int_{m(A)} v = \int_{m(A)} v_0 = S_{\rm{HEE}}^0 \, .
\eeq
\end{subequations}
Here,  $S_{\rm{HEE}}^0$ is the classical part of the HEE given by the area of the RT surface, and $ S_{\rm{bulk}}$ denotes the bulk entanglement entropy arising from quantum corrections. One can separate the quantum part of $v$ as,
\beq
 \int_{\partial \Sigma_A} v_q =  \int_{\Sigma_A} \nabla_{a} v_{q}^a =  S_{\rm{bulk}} =  \int_{\Sigma_A} s_{q}(x),
\eeq
with $\partial \Sigma_A =  A-m(A)$. The integrand $s_{q}(x)$ was termed the entanglement contour in \cite{Agon:2021tia}, but the authors emphasised that it need not be a positive definite quantity as it was originally introduced in \cite{Vidal:2014aal}.  However, we find that it is non-negative perturbatively in the first order. So, to avoid any confusion, we will call the quantity $s_{q}(x)$ the \emph{`bulk entanglement density'}, which gives rise to the bulk entanglement. The bulk entropy can also be written as the expectation value of the bulk modular Hamiltonian,
\beq
S_{\rm{bulk}} = \Big{<}\hat{H}_{\rm{bulk}} \Big{>} = \int_{\Sigma_A} \Big{<}T^{\rm{bulk}}_{ab} (x) \Big{>} \xi^a \varepsilon^b . \nonumber
\eeq
Here, $-n_a\wedge \varepsilon^a={\boldsymbol{\epsilon}}$, where ${\boldsymbol{\epsilon}}$ is the volume form for $d$-dimensional spacetime. From the above definition, it is clear that the divergenceless condition should be modified to incorporate the quantum part of the bit threads, and the modified divergence relation is given by,
\beq\label{modifieddivergence}
\nabla_{a} v_{q}^a =   s_{q}(x) = \Big{<}T^{\rm{bulk}}_{ab} (x) \Big{>} \xi^a n^b .
\eeq

\subsubsection*{Quantum corrections and energy condition}
The canonical bit threads prescription can be extended to include quantum corrections as well. First, let us examine the classical case with matter. Recall the fundamental identity of the CPS formalism,
\beq\label{fundamental_identity}
\bom(\delta \Phi,  \mathcal{L}_\xi \Phi,  \Phi) =  \mathrm{d} \mathbf{k}_\xi +  \xi^{a}\delta \bos{C}_a,
\eeq
where, $\delta \bos{C}_a$ is related to the constraints of the theory. In presence of matter, $\delta \bos{C}_a = -\delta T_{ab} \varepsilon^b$. So even if the symplectic current vanishes (in the presence of a Killing vector $\xi$), $\mathbf{k}_\xi$ does not remain a closed form: $\mathrm{d} \mathbf{k}_\xi = -\xi^{a}\delta \bos{C}_a$. The vector $v^a_{th}$ constructed from this codim-2 form will also not be divergenceless.\footnote{ It implies that, similar to the higher curvature gravity case \cite{Harper:2018sdd},  the divergence relation of bit threads should be modified in the presence of classical matter as well.} In the semiclassical regime, we consider the matter to be composed of quantum fields. Then the constraint becomes $\delta \bos{C}_a = -\langle \delta T_{ab} \rangle \varepsilon^b$. For exact symmetries, the identity \eqref{fundamental_identity} reduces to,
\beq
\mathrm{d} \mathbf{k}_\xi =- \xi^{a}\delta \bos{C}_a = \langle \delta T_{ab} \rangle \xi^a \varepsilon^b .
\eeq
Now, using the map \eqref{vk}, we can construct a vector field $\delta v^a$ which satisfies the following relation:
\beq\label{modifieddivergence2}
\nabla_a \delta v^a = \langle \delta T_{ab} \rangle \xi^a \varepsilon^b.
\eeq
 Here, $\dv = \dv_{0} + \dv_{q}$. We have changed the notation from $v^a_{th}$ to $\delta v^a$ to emphasise that the quantum bit threads in discussion are perturbative in nature. The divergence is also related to the perturbed bulk entanglement density as: $\nabla_a \delta v^a =\delta s_{q}(x)$. Integrating both sides of the above equation over the region $\Sigma_A$, we find
\begin{align}
	&\int_{\Sigma_A} \nabla_a \delta v^a = \int_{\Sigma_A} \langle \delta T_{ab} \rangle \xi^a \varepsilon^b \nonumber \\
	\Rightarrow &\int_A \delta v = \int_{m(A)} \delta v + \int_{\Sigma_A} \langle \delta T_{ab} \rangle \xi^a \varepsilon^b
\end{align}
This correctly captures the quantum corrections in this regime. Therefore, the positivity inequality \eqref{positivity} of relative entropy can be rewritten in terms of the bit threads vector field as follows
\beq
\int_{\Sigma_A} \mathrm{d} \mathbf{k}_\xi \; \ge 0 ~~ \Rightarrow ~ \int_{\Sigma_A} \nabla_a \dv^a \; \ge 0 \, .
\eeq
 Using the modified divergence relation \eqref{modifieddivergence2} we find,
\begin{align}
	& \ \int_{\Sigma_A} \delta s_q (x) \; \ge 0 \nonumber \\
	\Rightarrow &  \int_{\Sigma_A} \Big{<}\delta T^{\rm{bulk}}_{ab} (x) \Big{>} \xi^a \varepsilon^b  \; \ge 0\,  .
\end{align}
This relation is true for any arbitrary size of the boundary region $A$. So the above inequality is true for the integrand itself, which gives us the energy condition
\beq
\Big{<}\delta T^{\rm{bulk}}_{ab} (x) \Big{>} \xi^a n^b  \; \ge 0 \, .
\eeq
We find that this energy condition is equivalent to the dominant energy condition (DEC).\footnote{ To check it, we use the physical form of the DEC for perturbative matter
\beq
\delta T_{ab} \xi^a X^b  \; \ge 0 \nonumber \,,
\eeq
with $F^a =-\delta T^{a}_{\;\;b} \xi^b $ being a future pointing causal vector, and $\xi^a X_a < 0$. For our case, let $X^a=n^a$ be a unit normal timelike vector. Then it satisfies the condition $\xi^a n_a < 0$. Also, as $F^a$ and $n^a$ both are future-pointing and causal, they satisfy
\beq
	 F^a n_a \le 0 \qquad \Rightarrow \; -\delta T^{a}_{\;\;b} \xi^b n_a \le 0  \qquad
	\Rightarrow ~~ \delta T_{ab} \xi^b n^a \ge 0	\, \nonumber,
\eeq
which is nothing but the classical version of the energy condition that we have obtained.} So we conclude that if the perturbative bulk stress-energy tensor satisfies the DEC, then the bulk entanglement density should be positive: $\delta s_q (x) \ge 0$. It subsequently implies the positivity of the relative entropy. 

\section{Bit threads in AdS-Vaidya}\label{sec:dynamical}
In this section, we attempt to apply a similar method to the one employed in the previous sections to certain dynamical spacetimes.
 
 Consider those dynamical spacetimes that can be mapped to static spacetimes via some conformal transformations. It has been shown that for such cases, the conformal relation plays an important role in deriving the thermodynamics of the dynamical black holes \cite{Nielsen:2017hxt}. Another example of this kind of geometry is the AdS-Vaidya spacetime.
Here, we will look into the AdS-Vaidya spacetime in 3 dimensions. The metric of this spacetime is
\beq \label{ads-vaidya}
ds^2 = - \left( r^2 - m(w) \right) dw^2 + 2dw dr + r^2 dx^2,
\eeq 
where $m(w)$ is the mass function given by
\beq
 m(w) = M \,\theta(w) ~~ \rm{with} ~~ \text{$\theta(w)$}=\begin{cases}
    1, & \text{for $w \ge 0$}.\\
    0, & \text{for $w < 0$}.
  \end{cases}
\eeq
Here, we consider the mass $M$ to be used for parametric variations. The conformal Killing vector here is: $\xi = (\partial_w ,0,0)$. Using \eqref{keh}, we find 
\beq\label{kadsvaidya}
16 \pi G_N k ^{wr}_\xi = - \frac{3 \hdel M }{4 \left(r^2-m(w)\right)^2}\left(r^3 \theta (w)-M \delta_{DD} (w) \theta (w)\right)+\frac{\hdel M \delta_{DD} (w)}{r^2-m(w)}-\frac{\hdel M (2 r \theta (w))}{r^2 -m(w)}, \nonumber
\eeq
where $\delta_{DD} (.)$ represents the Dirac delta function. We can find a perturbative flow by taking the Hodge dual of this form and contracting with the hypersurface metric. There exists a $\bos{\mathit{j}}_\xi$ such that $\mathbf{k}_\xi = \pm \hdel \bos{\mathit{j}}_\xi$ given by,
\begin{align}\label{j_vaidya}
16 \pi G_N \mathit{j}^{wr}_\xi = &-\frac{3 r^2 \delta_{DD} (w)}{4 m(w)-4 r^2 \theta (w)}-\frac{\delta_{DD} (w) \log \left(r^2-m (w)\right)}{4 \theta (w)} \nonumber\\
&+\frac{r^3 (11 \theta (w)-8)-8 r \left(r^2-m (w)\right) \log \left(r^2-m (w)\right)}{4 \theta (w) \left(M-r^2\right)} \,. 
\end{align} 
Now, to find the vector $v^a_{cps}$, we need to specify a Cauchy surface. On the other hand, as the spacetime is dynamic, the extremal surface need not lie on a Cauchy surface. However, to apply our method, we will restrict ourselves to a constant time slice and examine the spacelike extremal surfaces. We will adopt an approach similar to that of \cite{Caginalp:2019mgu}. First, we rewrite the metric \eqref{ads-vaidya} in terms of $t$ and $r$ coordinates:
\beq\label{ads-vaidya2}
ds^2 = - f(r) dt^2 + \dfrac{dr^2}{f(r)} + r^2 dx^2 \, , 
\eeq
where, $f(r)=r^2$ in AdS and $f(r)=(r^2 -M)$ in BTZ. In the AdS phase, the current form \eqref{j_vaidya} reduces to
\beq
 \mathit{j}^{tr}_\xi = \dfrac{1}{16 \pi G_N } \dfrac{\mathfrak{C}}{r},
\eeq
where $\mathfrak{C}$ is an overall constant that can be fixed by imposing the boundary conditions. The above expression is quite similar to the pure AdS case, which suggests that the codim-2 current we have obtained is correct. Using the map \eqref{vj}, we can find a divergenceless vector field,
\beq\label{v_cps-ads_vaidya}
 v^a_{cps} = \left(- v^r_{cps} , v^x_{cps} \right) =- \dfrac{1}{4 G_N} (\mathfrak{C}r,0) .
\eeq
To check whether it gives rise to a flow or not depends on whether the vector field is proportional to the unit normal of the extremal surface corresponding to a boundary subregion. We can use the geodesic equation to find the minimal surface (spacelike geodesic) for a boundary subregion centred at $x=0$. The RT surface equation then reads:
\beq\label{rt1-ads_vaidya}
r = \dfrac{p_x}{\sqrt{x^2 p_x^2 -1 }}\, , \nonumber
\eeq
where $p_x$ is a conserved quantity along the RT surface. We can rearrange the above equation a little to obtain a nice form as,
\beq\label{rt1.2-ads_vaidya}
 x^2 + \dfrac{1}{r^2} = \dfrac{1}{p_x^2}\, .
\eeq
We can find the unit normal to this surface as given by,
\beq\label{normal_ads-vaidya}
n^a = (n^r,n^x) = \left( -p_x , \dfrac{x p_x}{r} \right) .
\eeq
Now, comparing the vector \eqref{v_cps-ads_vaidya} with the above unit normal of RT surface \eqref{rt1.2-ads_vaidya}, one can easily check that $v^a_{cps}|_{RT}\neq \mathrm{C}n^a$. So, we need to add $v^a_\ed$ to make the vector field a bit thread flow. We solve the differential equation $\nabla_a v^a_\ed = \nabla_a \nabla^a \varphi = 0$ in the background \eqref{ads-vaidya2} with proper boundary conditions, and find
\beq
\varphi = \dfrac{1}{4 G_N} \left[ \dfrac{x}{p_x} \sqrt{r^2 - p_x^2}\left( \mathfrak{C}r + p_x \right) +\dfrac{p_x}{r} - \dfrac{1}{2} \mathfrak{C}r^2 x^2 \right].
\eeq 
The total vector field is given by,
\beq
v^a =- \dfrac{1}{4 G_N}  \left( \mathfrak{C}r - p_x - \dfrac{\mathfrak{C}p_x}{\sqrt{x^2 p_x^2 -1}}, \left( \mathfrak{C}r + p_x \right) \dfrac{\sqrt{r^2-p_x^2}}{r^2 p_x} -\mathfrak{C}x  \right).
\eeq
It reduces to the unit normal \eqref{normal_ads-vaidya} at the RT surface \eqref{rt1.2-ads_vaidya}, which confirms that the above vector field is a bit thread vector field. One can also go to the Poincar\'{e} coordinates and find a flow, for which we have already shown that the vector field obtained from the CPS is not enough. We need to add the term $v^a_\ed$ so that the total vector field becomes a bit thread configuration. 

 Now, we look into the BTZ phase ($w\geq 0$). In the $(t,r,x)$ coordinates, the form \eqref{j_vaidya} becomes \eqref{j-btz},
\beq
  j_\xi^{tr} = \dfrac{1}{16\pi G_N r} M, \nonumber
\eeq
with AdS radius $L=1$. The vector obtained from this form is given by \eqref{v_cps-btz},
\beq\label{v-vaidya}
  v_{cps}^a = \left(- v_{cps}^r,v_{cps}^x \right) = -\dfrac{1}{4 G_N}\left( \dfrac{ \sqrt{M}}{ r} \sqrt{r^2 - M}, 0 \right). 
\eeq
We need to look at whether the vector field becomes proportional to the unit normal at the RT surface. We can find the RT surface equation using the geodesics entirely in the BTZ phase and using the metric \eqref{ads-vaidya2}. It is given by
\beq\label{rt2-ads_vaidya}
x =  F\left(\sin ^{-1}(r)\,\bigg|\, \frac{1}{p_x^2}\right)-E\left(\sin ^{-1}(r)\,\bigg|\, \frac{1}{p_x^2}\right) - \dfrac{\sqrt{(r^2-p_x^2)(r^2-M)}}{p_x \,r} \, ,
\eeq
where $F(\mathsf{k}|\mathsf{m})$ is the incomplete elliptic integral of the first kind and $E(\mathsf{k}|\mathsf{m})$ is the incomplete elliptic integral of the second kind.\footnote{The incomplete elliptic integrals of the first and second kind are given by \cite{abramowitz1965handbook},
\beq
 F(\mathsf{k}|\mathsf{m}) = \int_0^{\sin{\mathsf{k}}} \left[ (1-t^2)(1-\mathsf{m}\,t^2)\right]^{-1/2}\, ; \quad E(\mathsf{k}|\mathsf{m}) = \int_0^{\sin{\mathsf{k}}} (1-t^2)^{-1/2}(1-\mathsf{m}\,t^2)^{1/2}. \nonumber
\eeq} The unit normal to this surface is given by,
\beq\label{normal_ads-vaidya2}
 n^a = (n^r,n^x) = \dfrac{1}{r^2} \left(-r p_x \sqrt{r^2 -M}, \sqrt{r^2 -p_x^2} \right) .
\eeq
From the above expression and \eqref{v-vaidya}, it is clear that the resultant vector field $v^a_{cps}$ is not proportional to the normal $n^a$ at the RT surface \eqref{rt2-ads_vaidya}. So, we need to add the ambiguity field $v_{\ed}^a$ to make it a flow for a valid bit thread configuration. One can find this $v_{\ed}^a$ by using the method of characteristics with proper boundary conditions, as we did for the pure AdS phase of the AdS-Vaidya geometry. 

So we see that although the AdS$_3$-Vaidya spacetime does not have a Killing vector, the presence of a conformal Killing vector can provide us with the codim-2 forms and the corresponding vector field. This is surprising in a way that even in certain dynamical spacetimes, we can define (perturbative) bit thread flows using conformal Killing vectors rather than genuine Killing vectors. It extends the bit thread formalism beyond static spacetimes, though finding and interpreting the complete non-perturbative flow in general remains an open question.

\section{Conclusion and Discussion}\label{sec:discussion}
In this work, we have explored the bit thread formalism and its broader applicability to geometric entropy concepts beyond the conventional HEE. Our study has centred on several key themes: formulating methods based on CPS formalism to obtain bit thread configurations, their connections to Wald and differential entropy, a reformulation of thermodynamic first laws in terms of flow dynamics, and preliminary steps toward accommodating quantum corrections and time dependence.

We attempt to construct a bit thread configuration using quantities from the CPS formalism. We found that although it naturally produces a divergenceless vector field $v_{cps}^a$, it generally does not meet the norm bound and saturation conditions required of HEE threads $ v^a $.  A gauge freedom $ v_\ed^a$ needed to satisfy the constraints of bit threads can be obtained by solving for a harmonic function in $3$-dimensional spacetime. We showed that the bit threads vector field and $v^a_{cps}$ are in the same equivalence class $v^a \sim v_{cps}^a$. This is one of the main results of our work. In spacetimes where the RT surface becomes a horizon, the vector $v^a_{cps}$ is same as the bit thread vector field. Then, applying suitable coordinate transformations to it results in a bit thread configuration in the original metric. We used this method to obtain a bit thread configuration for Poincar\'{e} AdS in section \ref{sec:btcps}. This method does not require any prior knowledge of the RT surface and its unit normal; however, surprisingly, we find that the resultant thread configuration is same as the geodesic bit threads \cite{Agon:2018lwq} up to an overall constant. These alternative ways for bit thread construction may be more suitable for specific setups.

We have also extended this flow prescription to rewrite other entropies in the bit thread language. First, we related the horizon entropy to this flow picture. In section \ref{sec:extension}, we showed how a bit thread-like vector can be found for AdS black holes, which gives the Wald entropy once integrated over the bifurcation surface. Another entropy that can be expressed in terms of threads is the differential entropy for certain spacetimes, which is also defined for non-holographic spacetimes.

The CPS approach also inspired a new way to view thermodynamic first laws (Section~\ref{sec:first_laws}). By utilising the Hodge dual of the current codim-2 form $\mathbf{k}_{\xi} $, we introduced the \emph{thermodynamic variation flow} \( v_{th}^a = g^{ab} (\star \mathbf{k}_{\xi})_b \), which captures the flux balances associated with entropy changes. These balance equations provide a geometric visualisation of how entropy variations arise from the bulk symplectic current, which is relevant for both black hole thermodynamics and the entanglement entropy first law. For differential entropy, we find that the first law relates the Lorentzian and Riemannian (perturbative) bit threads.

Finally, our preliminary exploration into quantum corrections (Section~\ref{sec:quantum}) and dynamical spacetimes (Section~\ref{sec:dynamical}) may provide new directions for future work. Extending bit threads into the quantum regime requires modifying the norm bound or divergence conditions to account for bulk entanglement, potentially connecting to the generalised second law and quantum energy conditions. Time-dependent settings necessitate working with closed \( (d{-}1) \)-forms and pose significant challenges in constructing thread configurations in the absence of symmetries.

Here are some questions and future research directions:
\begin{itemize} \itemsep.8em

\item[]\textsl{\underline{Higher curvature gravity:}} The bit threads picture has been explored in higher curvature gravities in general \cite{Harper:2018sdd}, but an explicit construction for some particular theory is still missing. The method used here to construct a flow using the CPS formalism can be extended to these higher-derivative gravities very naturally, as the CPS formalism is applicable to more general formulations of gravity \cite{Iyer:1994ys}. In \cite{Harper:2018sdd}, the authors showed how the norm bound changes for higher curvature gravity, while the vector field remains divergenceless. This property comes naturally following the CPS formalism, and one should be able to construct a bit thread configuration by imposing the modified norm bound. It can be further extended to the cubic gravitational theories to possibly address the issue of `splitting' as discussed in \cite{Caceres:2020jrf}.

\item[]\textsl{\underline{Towards covariant bit threads:}} The notion of covariant bit threads has been explored in \cite{Headrick:2022nbe}. Their exploration is mainly focused on the definition and general properties of the covariant bit threads. Unlike the Riemannian ones, the covariant threads live in the whole spacetime, and their flux is maximum at the HRT surface of the corresponding boundary region. The paper showed that we can find a covariant flow from a closed $(d-1)$-form obeying the required norm bound. If one wants to obtain these threads from CPS formalism, the Noether current \eqref{noethercurrent} is a promising candidate, which is a $(d-1)$-form residing in the spacetime. Conservation ensures that the Noether current is a closed form (on-shell). So one can find a divergenceless vector field by taking the Hodge dual of the Noether current,
\beq
 V^{\mu}_{cov} =g^{\mu \nu} (\ast \boldsymbol{\mathrm{J}}_{\xi})_\nu \, .
\eeq
We hope that this vector field can be related to the `V-threads' presented in \cite{Headrick:2022nbe}, but the exact relation needs further investigation. Identifying these threads explicitly will enhance our understanding of the relationship between entanglement and geometry. This method may also be helpful to extend the bulk reconstruction in the covariant setting, similar to the Riemannian case, as done in \cite{Agon:2020mvu}.

\item[]\textsl{\underline{Quantum energy conditions:}} One interesting direction would be to derive the quantum focusing conjecture \cite{Bousso:2015mna} and corresponding quantum null energy conditions \cite{Bousso:2015wca} using this flow-based prescription. It will also help to understand the covariant entropy bound and quantum Bousso bound through the bit threads perspective, and may shed new light on the underlying structure of these quantities.

\item[]\textsl{\underline{Generalised entanglement wedge:}} The construction of a flow from CPS formalism is also applicable to general gravitational spacetimes. Recently, there has been a proposal of generalised entanglement wedge (GEW) in general gravitational spacetimes \cite{Bousso:2022hlz, Bousso:2023sya}, which states that for any gravitational bulk region $\mathtt{b}$, there is a GEW $\mathtt{E(b)}$ ($\supset \mathtt{b}$) on a static Cauchy surface. Then the fine-grained entropy of the region $\mathtt{b}$ is given by the generalised entropy of the GEW $S_{gen}(\mathtt{E(b)})$. The authors of \cite{Du:2024xoz} extended the bit thread formulation for the GEW proposal in general gravitational spacetimes. The construction of bit threads using the CPS formalism provides a canonical way to extend the flow prescription to this GEW proposal.

\item[] \textsl{\underline{Timelike entanglement entropy:}} It would be interesting to find a flow-based picture for the timelike entanglement entropy \cite{Doi:2023zaf}. One possible approach is to use the Wick-rotated Euclidean metric in AdS and apply the method described in section \ref{sec:btcps}. It may also be related to the pseudo entropy \cite{Nakata:2020luh}, which is more general than the timelike EE.

\item[] \textsl{\underline{Connection to geometric modular flows:}} In CFTs, entanglement entropy of a subregion can be obtained by using the notion of modular geometric flows \cite{Caminiti:2025hjq}. In the holographic context, the same can be obtained by using the bit threads. A possible connection has already been mentioned in \cite{Caminiti:2025hjq}. The difference between these two formalisms is that the geometric modular flow is uniquely determined, whereas the configuration of bit threads is highly non-unique. Apparently, the bit threads and the geometric modular flows are normal to each other at the boundary subregion. If we consider the case in the AdS-Rindler wedge, the CPS method provides a canonical thread configuration. It will be worth exploring the connection between these two kinds of flows in future. 

\end{itemize}

\section*{Acknowledgments}
We would like to thank Debajyoti Sarkar, Avik Chakraborty, and Mrityunjay Nath for helpful discussions on various stages of this work. We are also grateful to the anonymous referee for his/her constructive review and helpful comments. PKD acknowledges the financial support from the DST-INSPIRE fellowship (IF200253), Department of Science and Technology, Govt. of India. This work is also supported by DST-FIST initiative (SR/FST/LS-I/2020/621) and (GrantSR/FST/PSI-225/2016).

\appendix
\renewcommand{\theequation}{\thesection.\arabic{equation}}

\section{Bit threads: a brief review}\label{appsec:bithtreads}

Here we will briefly review the notion of bit threads proposed by Headrick and Freedman \cite{Freedman:2016zud}. We will start from the continuation version of the well-known max-flow min-cut theorem, which originated in network theory.

Given an oriented $d$-dimensional Riemannian manifold with boundary, a \emph{`flow'} is a vector field $\upsilon$ satisfying,
\begin{equation} \label{flow}
	\nabla_{a} \upsilon^{a} = 0\,,\qquad|\upsilon|\le \mathrm{C}\,,
\end{equation}
where, $\mathrm{C}$ is a positive quantity. A \emph{`surface'} is defined as a $(d-1)$-dimensional oriented submanifold. Then, the flux of the flow through such a surface `$m$' can be expressed as
\begin{equation}\label{flux}
	\int_m \upsilon:=\int_m\sqrt{\mathfrak{h}}\,n_a \upsilon^a ,
\end{equation}
where, $\mathfrak{h}$ is the determinant of the induced metric on $m$ and $n^a$ is the unit normal vector. The divergenceless condition \eqref{flow} implies 
\begin{equation}
	m\sim A\quad\Rightarrow\quad\int_m \upsilon=\int_A \upsilon\,,
\end{equation}
where, $m\sim A$ denotes that the surfaces $m$ and $A$ are homologous to each other.
\begin{figure}[htbp]
 \centering
	\includegraphics[scale=0.03]{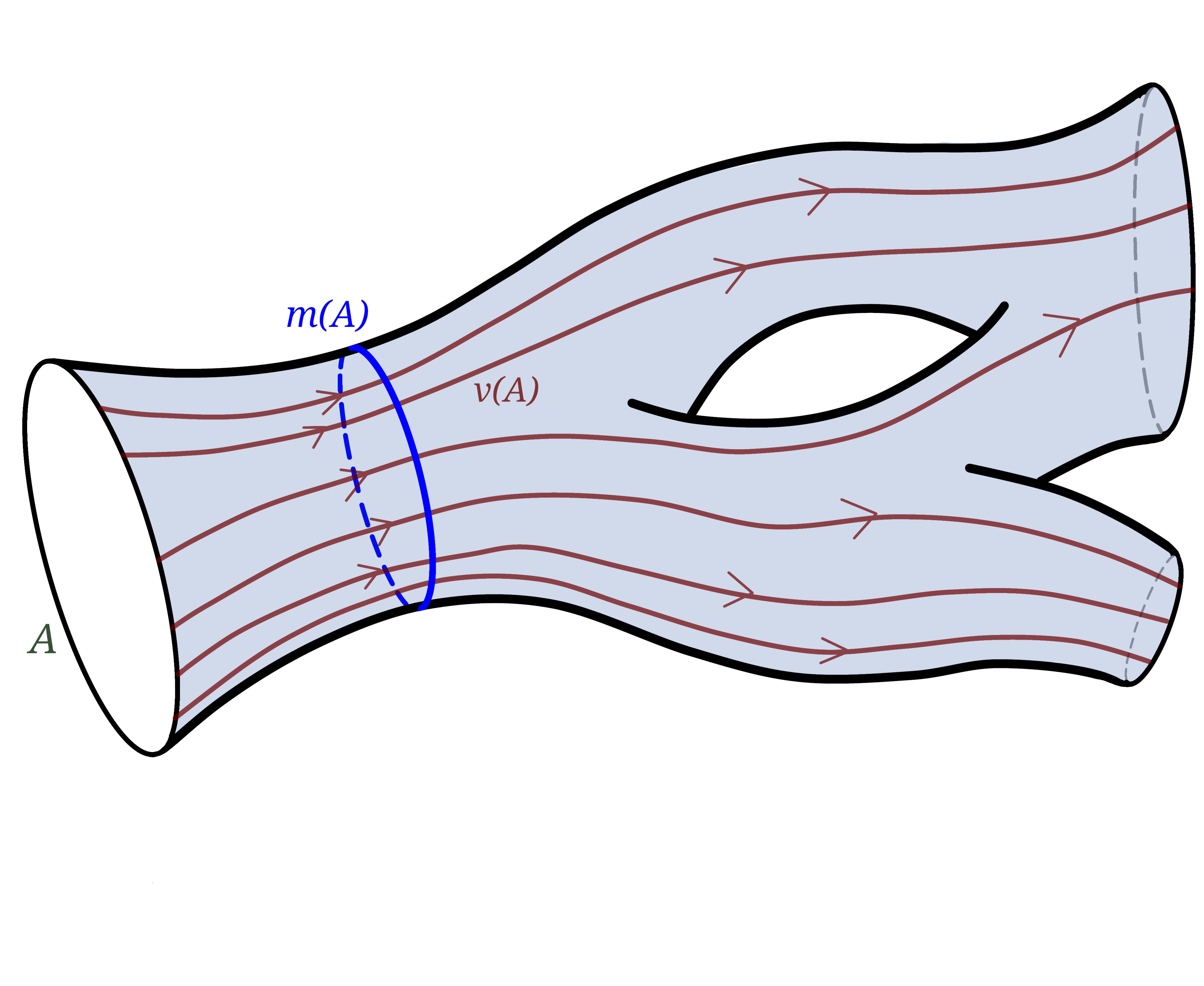}
	\caption{\textit{Schematic diagram of a Riemannian manifold showing a flow $\upsilon(A)$ and a surface $m(A)$ with minimal area, homologous to the boundary region $A$. The Riemannian MFMC tells that the flux of the flow (flow lines shown in brown) is proportional to the area of the surface $m(A)$ (shown in blue). This minimal surface acts as the bottleneck for the flow.}}\label{fig:mf}
\end{figure}

The norm bound $|\upsilon|\le \mathrm{C}$ implies that the flux is bounded by the area of $m$:
\begin{equation} \label{ineq}
	\int_m \upsilon \le \mathrm{C} \int_m \sqrt{h} = \mathrm{C}\; \cA(m)\,,
\end{equation}
where $\cA(m)$ denotes the area of $m$. Maximising on one side over all the flows $\upsilon$ and minimising on the other over all surfaces homologous to $A$ yields
\begin{equation} \label{flineq}
	\max_{\upsilon} \int_A \upsilon \le \mathrm{C} \min_{m\sim A}\; \cA(m)\,.
\end{equation}
The Riemannian max-flow min-cut (MFMC) theorem \cite{Freedman:2016zud, Headrick:2017ucz} states that the above inequality is saturated:
\begin{equation} \label{mfmc}
	\max_{\upsilon} \int_A \upsilon = \mathrm{C} \min_{m\sim A}\; \cA(m)\,.
\end{equation}
The inequalities \eqref{ineq} provide an upper bound for the flux of any flow via the area of a minimal surface $m(A)$ (homologous to $A$), which is the bottleneck  (as shown in figure \ref{fig:mf}).\footnote{We thank Ayushi Suman for the figure.} The MFMC theorem \eqref{mfmc} implies that $n_a \upsilon^a=\mathrm{C}$, and hence $\upsilon^a=\mathrm{C}n^a$, everywhere on $m(A)$.

In the holographic context, one can use the above MFMC theorem to rewrite the RT formula for entanglement entropy. In a bulk theory, the Riemannian manifold is a constant-time slice, and $A$ is a boundary subregion in conformal field theory. Then, with the help of equation \eqref{mfmc}, and by setting $\mathrm{C} = 1/4 G_{N}$, the RT formula can be reexpressed as,
\begin{equation}\label{defbt}
	S_{\rm{HEE}}(A) = \max_{\upsilon} \int_A \upsilon\,.
\end{equation}

\section{Covariant phase space formalism}\label{sec:cps}
The covariant phase space formalism \cite{Crnkovic:1986ex, Iyer:1994ys, Wald:1993nt} is an approach to study Hamiltonian mechanics in which one constructs a phase space by taking the set of solutions to the equations of motion instead of the generalised coordinates and their conjugates. To study the entropy and thermodynamics of black holes, people mainly use the Iyer-Wald formalism \cite{Iyer:1994ys}. There have been many modifications and generalisations to this formalism. One such example is \cite{Harlow:2019yfa}, where the authors considered the boundary Lagrangian separately. However, the entropy obtained from both of these formalisms is the same. Here, we will briefly discuss both of these formalisms, the Wald entropy and the well-known first law of black hole thermodynamics.

\subsection{Iyer-Wald Formalism}

In CPS, the phase space is considered to be a manifold $\cP$ with a symplectic $2$-form $\Omega$, and the spacetime is a $d$-dimensional manifold $\cM$. Then, from the symplectic form, one can define the Hamiltonian. The action for a general theory is
\begin{equation}
	\mathrm{S}[\Phi] = \int_{\mathcal{M}} \boldsymbol{\mathrm{L}}(\Phi),
\end{equation}
where, $\boldsymbol{\mathrm{L}}$ is the Lagrangian $d$-form for any $d$-dimensional theory, and $\Phi$ denotes all the dynamical fields collectively. Taking a variation of this action leads to
\begin{equation}
	\delta \boldsymbol{\mathrm{L}}[\Phi] = \boldsymbol{\mathrm{E}}_{\Phi}  \delta \Phi + \mathrm{d} \mathbf{\Theta}_{\rm{IW}} \left( \delta \Phi, \Phi \right),
\end{equation}
where $\delta \Phi$ represents a generic field perturbation and provides the basis of (tangent space of) the phase space $\mathcal{P}$. The equations of motion are given by the set of equations $\boldsymbol{\mathrm{E}}_{\Phi} = 0$. The quantity $\mathbf{\Theta}_{\rm{IW}}$ is called the symplectic potential. It is a $(d-1)$-form in spacetime $\mathcal{M}$ and one form on the phase space, i.e., $\mathbf{\Theta}_{\rm{IW}}$ is a $(d-1;1)$-form. Then, one can define the quantity called (pre)symplectic current as
\begin{equation}\label{sympcurrent}
	\bom_{\rm{IW}}(\delta_{1} \Phi, \delta_{2} \Phi,  \Phi) = \delta_{1} \mathbf{\Theta}_{\rm{IW}}(\delta_{2} \Phi, \Phi) - \delta_{2} \mathbf{\Theta}_{\rm{IW}} (\delta_{1} \Phi,  \Phi).
\end{equation}
Here, $\delta_{1} \Phi$ and $\delta_{2} \Phi$ are two arbitrary field perturbations, `$\mathrm{d}$' denotes the exterior derivative over the spacetime $\mathcal{M}$ and `$\delta$' represents the exterior derivative over the phase space manifold $\mathcal{P}$. Using this notation, it is evident that $\bom_{\rm{IW}}$ is a $2$-form on the phase space, or more compactly, is a $(d-1;2)$-form. A symplectic form then can be constructed by integrating the symplectic current over a codimension-one spacelike surface $\Sigma$,
\begin{equation}
	\mathbf{\Omega}_{\rm{IW}} (\delta_{1} \Phi, \delta_{2} \Phi,  \Phi) = \int_{\Sigma} \bom_{\rm{IW}}(\delta_{1} \Phi, \delta_{2} \Phi,  \Phi).
\end{equation}
This $\mathbf{\Omega}_{\rm{IW}}$ is non-degenerate, and it does not depend on the choice of the Cauchy surface $\Sigma$. If $\mathcal{F}$ is the set of solutions of the equations of motion, then $ (\mathcal{F}; \mathbf{\Omega})$ constitutes a well-defined phase space denoted by $\cP$.

Now, given a transformation generated by $\xi$, one can define different quantities like Noether charge, Noether current, and the Hamiltonian (or, more generally, a conserved charge). For diffeomorphism invariant theories, covariance tells that $\delta_\xi \Phi = \mathcal{L}_\xi \Phi$, where $\xi$ is the diffeomorphism generator. So the (pre)-symplectic current becomes,
\begin{equation}
	\bom_{\rm{IW}}(\delta \Phi, \mathcal{L}_\xi \Phi,  \Phi) = \delta \mathbf{\Theta}_{\rm{IW}}(\mathcal{L}_\xi \Phi, \Phi) - \mathcal{L}_\xi \mathbf{\Theta}_{\rm{IW}} (\delta \Phi,  \Phi).
\end{equation}
One can now introduce the Noether current $\boldsymbol{\mathrm{J}}$ as,
\begin{equation}\label{noethercurrent}
	\boldsymbol{\mathrm{J}}_{\xi} \equiv \mathbf{\Theta}_{\rm{IW}} (\mathcal{L}_\xi \Phi, \Phi) - \xi .\boldsymbol{ \mathrm{L}} ~ .
\end{equation}
A simple check shows that this current is a closed form, and locally, it can be made an exact form, which leads to the definition of the Noether charge, 
\begin{equation}\label{noethercharge}
	\boldsymbol{\mathrm{J}}_{\xi} = \mathrm{d} \boldsymbol{ \mathrm{Q}}_{\xi}.
\end{equation}
The Hamiltonian can then be written as,
\begin{equation}\label{hamiltonianIW}
	\delta H_\xi = \mathbf{\Omega}_{\rm{IW}}(\delta \Phi, \mathcal{L}_\xi  \Phi, \Phi) = \int_\Sigma \bom_{\rm{IW}}.
\end{equation}
The above expression can be written in terms of the Noether current and Noether charge as well. Now, we will talk about the recently modified version of the covariant phase space with boundaries.

\subsection{CPS with Boundaries}

Recently, Harlow and Wu generalised the CPS formalism by adding a boundary term, and the result is the \textit{covariant phase space formalism with boundaries} (CPSB) \cite{Harlow:2019yfa}. In this kind of formulation, the action is written as,
\begin{equation}\label{actioncpsb}
	\mathrm{S}[\Phi] = \int_{\mathcal{M}} \boldsymbol{\mathrm{L}} \big(\Phi, \Xi \big) + \int_{\partial \mathcal{M}} \boldsymbol{{\ell}} \big(\Phi, \Xi \big)
\end{equation}
with boundary conditions for $\Phi^{i} (x)$ on the spatial boundary $\Gamma$. Here, $\Phi$ is dynamical and $\Xi$ is the background field. Then one imposes that the action is well posed, i.e. $\Phi$ obeys EOM iff $\mathrm{S}$ is stationary (up to some boundary terms in future and past slices). Similar to the IW formalism, here also the Lagrangian is considered to be a $d$-form (for $d$-dimensional spacetime manifold) instead of a scalar. Therefore, the variation of the Lagrangian can be written as
\begin{equation}
	\delta \boldsymbol{ \mathrm{L}} = \boldsymbol{\mathrm{E}}_{i} \delta \Phi^{i} + \mathrm{d } \mathbf{\Theta}_{\rm{HW}}
\end{equation}
where, `$\mathrm{d}$' is the exterior derivative in space-time, $\mathbf{\Theta}_{\rm{HW}}$ is called symplectic potential. Thus, the variation of the action is given by
\begin{equation}
	\delta \mathrm{S} = \int_{\mathcal{M}} \boldsymbol{\mathrm{E}}_{i} \delta \Phi^{i} + \int_{\Sigma_{\pm}} \big( \mathbf{\Theta}_{\rm{HW}} + \delta \boldsymbol{\ell} \big) + \int_{\Gamma} \big( \mathbf{\Theta}_{\rm{HW}} + \delta \boldsymbol{\ell} \big),
\end{equation}
where, the boundary of the manifold $\mathcal{M}$ is: $\partial\mathcal{M}= \Gamma \cup \Sigma_{-} \cup \Sigma_{+}$. $\Sigma_{\pm}$ are the future (past) boundaries and $\Gamma$ is the spatial boundary. As mentioned earlier, the action must be stationary. The stationarity of $\mathrm{S}$ implies $\delta \mathrm{S} = 0$. This condition results in two important equations:
\begin{equation}
	\boldsymbol{\mathrm{E}}_i = 0  \quad (Equations\; of\; Motion)
\end{equation}
and the other one is the non-trivial one,
\begin{equation}
	\big( \mathbf{\Theta}_{\rm{HW}} + \delta \boldsymbol{\ell} \big)\mid_{\Gamma} = \mathrm{d} \boldsymbol{ \mathsf{C}}
\end{equation}
for any $(d-2)$-form $\boldsymbol{ \mathsf{C}}$. Now, one can define a quantity named \textit{pre-symplectic current}, $\bom_{\rm{HW}} \equiv \delta(\mathbf{\Theta}_{\rm{HW}} - \mathrm{d} \boldsymbol{\mathsf{C}})\mid_{\tilde{\mathcal{P}}}$ with certain properties. Integrating the pre-symplectic current over a Cauchy slice, one obtains the \textit{pre-symplectic form},
\begin{equation}
	\tilde{\mathbf{\Omega}}_{\rm{HW}} = \int_{\Sigma} \bom_{\rm{HW}}
\end{equation}
which is `closed' and independent of $\Sigma$. The pre-symplectic form is used to construct the Hamiltonian associated with this CPSB. Consider a diffeomorphism generator $\xi^{\mu}$ such that (i) $\xi^{\mu}$ preserves the boundary conditions, (ii) $\boldsymbol{ \mathrm{L}}$ and $\boldsymbol{{\ell}}$ are \textit{covariant} w.r.t $\xi^{\mu}$. Then, the Hamiltonian $H_{\xi}$ is defined such that,
\begin{equation} \label{hamiltonian}
	\delta H_{\xi} = - X_{\xi}.\tilde{\mathbf{\Omega}}_{\rm{HW}},
\end{equation}
where, $ X_{\xi} \equiv \int d^{d}\mathrm{x} \; \mathcal{L}_{\xi} \; \Phi^{i}(x)\dfrac{\delta}{\delta \Phi^{i}(x)}$. The Noether current is defined as,
\begin{equation}
	\boldsymbol{\mathrm{J}}_{\xi} \equiv X_{\xi}.\mathbf{\Theta}_{\rm{HW}} - \xi. \boldsymbol{ \mathrm{L}} .
\end{equation}
Simple calculation using the above three equations gives,
\begin{equation} \label{xom}
	-X_{\xi}.\tilde{\mathbf{\Omega}}_{\rm{HW}} = \delta \left(\int_{\Sigma} \boldsymbol{\mathrm{J}}_{\xi} + \int_{\partial \Sigma} \big(\xi.\boldsymbol{{\ell}} - X_{\xi}.\boldsymbol{ \mathsf{C}}\big) \right) .
\end{equation}
Evidently, from equations \eqref{hamiltonian} and \eqref{xom}, one obtains the Hamiltonian,
\begin{equation}
	H_{\xi} = \int_{\Sigma} \boldsymbol{\mathrm{J}}_{\xi} + \int_{\partial \Sigma} \big(\xi.\boldsymbol{{\ell}} - X_{\xi}.\boldsymbol{ \mathsf{C}} \big) .
\end{equation}
This is true for any kind of higher derivative theory. Some examples can be found in \cite{Harlow:2019yfa}. There are some special theories (like the theory of general relativity), where $\boldsymbol{ \mathrm{L}}$ is covariant under arbitrary $\xi^{\mu}$. This gives us  $\mathrm{d }\boldsymbol{\mathrm{J}}_{\xi} = 0$ and $\boldsymbol{\mathrm{J}}_{\xi}$ becomes an exact form. So, one can define a $(d-2)$-form $\boldsymbol{ \mathrm{Q}}_{\xi}$ such that
\begin{equation}
	\boldsymbol{\mathrm{J}}_{\xi} = \mathrm{d } \boldsymbol{ \mathrm{Q}}_{\xi},
\end{equation}
$\boldsymbol{ \mathrm{Q}}_{\xi}$ being the “Noether charge". For these special theories, the Hamiltonian gets simplified to,
\begin{equation}
	H_{\xi} = \int_{\partial \Sigma} \left( \boldsymbol{ \mathrm{Q}}_{\xi} + \xi.\boldsymbol{{\ell}} - X_{\xi}.\boldsymbol{ \mathsf{C}} \right) ,
\end{equation}
which shows that the Hamiltonian is a total boundary term. 

\subsection{Black Hole Thermodynamics}\label{sec:bhthermodynamics}
Consider a stationary rotating black hole with non-degenerate Killing vectors. Further, assume the black hole is axially symmetric, has a bifurcation, and has no electric charge. The horizon Killing vector is denoted by $\xi_H$. The Killing vector corresponding to the time evolution is $\xi_M  $, and the Killing vectors corresponding to the rotation is $\xi_{J}$. These Killing vectors are related by,
\begin{equation} \label{killingrelation}
	\xi_H = \xi_M - \Omega_{H} \xi_{J},
\end{equation}
where, $\Omega_{H} $ represents the angular velocity of the horizon. Redefine the generators
\begin{equation}\label{Killingredefn}
	\eta_M = \xi_M ; \qquad \eta_{J} = \xi_{J}; \qquad	\eta_S = \frac{2 \pi}{\kappa} \xi_H,
\end{equation}
where $\kappa$ is the surface gravity of the black hole. The obvious relation between these new generators is given by,
\begin{equation}\label{killingrelation2}
	\eta_S = \frac{2 \pi}{\kappa} \left( \eta_M - \Omega_{H}\eta_{J} \right)
\end{equation}
Now, one can define the conserved charges corresponding to the generators. The conserved charges for time  evolution and rotation are defined by integrating the codim-$2$ currents $\mathbf{k}_{\eta_M}$ and $\mathbf{k}_{\eta_J}$ at the spatial infinity,
\begin{subequations}\label{canonicaldefn}
\begin{equation} \label{cmass}
	\delta M := \oint_{\mathcal{S}_\infty} \mathbf{k}_{\eta_M} (\delta \Phi, \Phi) ; \\
\end{equation}
\begin{equation} \label{cangmom}
	\delta J := \oint_{\mathcal{S}_\infty} \mathbf{k}_{\eta_{J}} (\delta \Phi, \Phi),
\end{equation}
and the conserved charge corresponding to the horizon Killing vector by integrating $\mathbf{k}_{\eta_S}$ at the bifurcation surface,
\begin{equation} \label{centropy}
	\delta S := \oint_\mathcal{B} \mathbf{k}_{\eta_S} (\delta \Phi, \Phi).
\end{equation}
\end{subequations}
These conserved charges $\delta M$, $\delta J$, and $\delta S$ are the mass of the black hole, the angular momentum of the black hole horizon, and the entropy of the black hole, respectively. To obtain the first law of black hole thermodynamics using the conserved charges and their corresponding codim-2 currents, recall the fundamental identity \eqref{fundid},
\beq
\int_{\Sigma} \bom (\delta \Phi,  \mathcal{L}_{\xi_{\rm{H}}} \Phi,  \Phi)  = \int_{\Sigma} \mathrm{d} \mathbf{k}_\eta (\delta \Phi, \Phi).
\eeq
The Cauchy surface $\Sigma$ has boundary $\partial \Sigma = \mathcal{S}_{\infty} \cup (- \mathcal{B})$, where $\mathcal{B}$ is a surface (here, the bifurcation surface) homologous to $\mathcal{S}_{\infty}$ (the spatial infinity). As $\xi_{H}$ is a Killing vector, any variation along its flow should vanish, i.e., $\mathcal{L}_{\xi_{H}} \Phi = 0$, or more precisely,
\begin{equation}
	\int_{\Sigma} \bom (\delta \Phi,  \mathcal{L}_{\xi_{\rm{H}}} \Phi,  \Phi) = 0,
\end{equation}
which immediately leads to,
\beq
 \oint_{\mathcal{S}_\infty} \mathbf{k}_{\rm{\eta_M}} (\delta \Phi, \Phi) - \Omega_{\rm{H}} \oint_{\mathcal{S}_\infty} \mathbf{k}_{\rm{\eta_{J}}} (\delta \Phi, \Phi) - \frac{\kappa}{2 \pi} \oint_{\mathcal{B}} \mathbf{k}_{\rm{\eta_S}} (\delta \Phi, \Phi) = 0,
\eeq
where the relation \eqref{krelation} has been used. Now, using the definitions of the conserved charges \eqref{canonicaldefn}, 
\begin{equation} \label{1stlaw1}
	\delta M - \Omega_{\rm{H}} \delta J- \frac{\kappa}{2 \pi} \delta S = 0 \nonumber .
\end{equation}
The temperature of the horizon is defined as $T_{\rm{H}} = \frac{\kappa}{2 \pi}$. Substituting it, one obtains the well-known 1st law of black hole thermodynamics,
\beq \label{1stlaw2}
\delta M = T_{\rm{H}} \delta S + \Omega_{\rm{H}} \delta J .
\eeq

\section{Flow in \texorpdfstring{D1-brane}{} background}\label{sec:d1-brane}

  $D1$-brane is an object in $10D$ Type IIB supergravity. An effective theory for this system can be obtained in 3 dimensions by compactifying it on a 7-sphere. This effective $3D$ Einstein-dilaton system is governed by the action \cite{David:2009np},
 \beq
 \mathrm{S}_{\rm{ED}} = \dfrac{1}{16 \pi G_{N}} \int d^{3}x \sqrt{-g} \left[ R - \dfrac{\nu}{2} \partial_{\mu}\vartheta \partial^{\mu}\vartheta - \cV(\vartheta) \right],
 \eeq
 where, $\nu = \frac{16}{9}$; $\vartheta$ is the dilaton and its potential $\cV(\vartheta) = - \frac{24}{L^2} e^{\frac{4}{3} \vartheta}$. The $3D$ metric of the $D1$-brane is given by,
 \beq\label{3dmetricsc}
 ds^2_{\rm{D1}} = -\mathcal{C}_{T}(r)^2 dt^2 + \mathcal{C}_{R}(r)^2 dr^2 + \mathcal{C}_{X}(r)^2 dx^2,
 \eeq
with the metric components,
 \beq
 \mathcal{C}_{T}(r)^2 = \left( \frac{r}{L} \right)^{8} f ~;~ \mathcal{C}_{R}(r)^2 = \frac{1}{f} \left( \frac{r}{L} \right)^{2} ~;~ \mathcal{C}_{X}(r)^2 = \left( \frac{r}{L} \right)^{8}, 
 \eeq
 where $f = 1 - \dfrac{r_{0}^6}{r^6}$, $r_0$ being the non-extremal parameter. And the dilaton is given by,
 \beq
 \vartheta = - 3 \ln{\left(\dfrac{r}{L}\right)}.
 \eeq
The metric \eqref{3dmetricsc} contains two constants: the AdS radius $L$ and the non-extremal parameter $r_0$. We can use either of these two or both to do the parametric variations as discussed in section \ref{sec:btcps}. For simplicity, we will use the AdS radius $L$ as the variation parameter. Unlike previous examples, this solution comprises both the gravitational and a scalar (dilaton) sector. For the gravity part, we can use \eqref{keh} to find the non-zero component of the codim-2 current:
\beq
 k^{tr}_{EH}=- \frac{L \left(r^6 + 2 r_0^6\right)}{4 \pi G_{N} r^9} \hdel L ,
\eeq
where the suffix $EH$ denotes the Einstein-Hilbert part of the action. For the scalar sector, we can use the expression \cite{Hajian:2015xlp},
 \beq \label{ksc}
 k^{\alpha\beta}_{Sc} = \dfrac{1}{8 \pi G_{N}} \left[ \xi^{\beta} \tilde{\nu} \nabla^{\alpha} \vartheta \delta \vartheta - (\alpha \leftrightarrow \beta) \right],
 \eeq
 with $\tilde{\nu} = \frac{\nu}{4} = \frac{4}{9}$. Non-zero component of this current is given by,
 \beq
 k^{tr}_{Sc} = \frac{ L}{4 \pi G_{N} r^3}  \left(1-\frac{r^{6}_0}{r^6}\right) \hdel L.
 \eeq
 By adding these two different contributions, we get the total codim-2 current,
 \begin{align}
 	 k^{ab} & = k^{ab}_{EH} + k^{ab}_{Sc} \nonumber \\  \Rightarrow k^{tr} & = - \frac{3 L^2 r^{6}_0}{4 \pi G_{N} r^9} \hdel L .
 \end{align}
Now, we can find a $(d-2)$-form $\bos{\mathit{j}}_\xi$ such that $\mathbf{k}=-\hdel\bos{\mathit{j}}$ given as,
\beq
j^{tr} = \dfrac{1}{4 \pi G_{N}} \left( \dfrac{r_0^6 L^3}{r^9} \right) .
\eeq
Restricting this form on a constant time slice and taking the Hodge dual, we find a divergenceless vector field
\beq\label{v_d1}
 v^a_{cps} = \left(- v^r_{cps} ,  v^x_{cps} \right) = -\dfrac{1}{4 G_N} \left(\mathfrak{E}\dfrac{r_0^4 L}{r^5} \sqrt{1-\dfrac{r_0^6}{r^6}}, 0 \right).
\eeq
Now,  to check if it gives rise to bit threads, we need to see whether this vector is proportional to the unit normal of an RT surface corresponding to a boundary subregion. Let us first try to find the RT surface equation. Finding this surface for an interval (in the boundary) is akin to going away from the centre of the bulk, i.e., taking the limit $r \rightarrow \infty$ (or $r \gg r_0$). In this limit the metric \eqref{3dmetricsc} reduces to
\beq\label{metric_d1-brane}
 ds^2_{\rm{D1}} |_{r\gg r_0} = - \left( \frac{r}{L} \right)^{8} dt^2 + \left( \frac{r}{L} \right)^{2} dr^2 + \left( \frac{r}{L} \right)^{8} dx^2\,.
\eeq
We can try to simplify it further. Substituting $y = r^4$ with $L=1$ yields
\beq\label{d1_new}
 ds^2 = y^2 (-dt^2 +dx^2) + \dfrac{dy^2}{16 y}.
\eeq
The RT surface can be obtained by integrating over the following differential equation,
\beq\label{rt_diffeq}
 dx = \dfrac{y_0 dy}{4 y^{3/2}\sqrt{y^2-y^2_0}}\,,
\eeq
where $y_0$ is the deepest point of the RT surface in the bulk. Integrating both sides of the above equation gives us the RT surface equation
\beq\label{rt_d1-brane}
4 y^2_0 x^2y = (y^2 - y_0^2)\, {_2}F_1\left(\, \dfrac{1}{4}\, ,\, 1\,;\, \dfrac{3}{4}\, ;\, \dfrac{y^2}{y_0^2} \, \right)^2 .
\eeq
We can find the unit normal to this surface by using \eqref{rt_diffeq}, given by
\beq\label{normal_d1}
n^a= (n^y , n^x) = \dfrac{1}{y^2 \sqrt{y^2 - y^2_0}} \left(- 4 y_0 y^{3/2} ,  \sqrt{y^2 - y^2_0}  \right).
\eeq
 Clearly, the vector $\eqref{v_d1}$ does not match the above unit normal. One reason is that we evaluated the vector field in the original 3-dimensional metric \eqref{3dmetricsc}, whereas the unit normal or the RT surface itself is obtained in the limit $r>>r_0$. So, we need to find the vector in this limit. Proceeding the same way as before, we find the vector field obtained from CPS as
\beq\label{v_d1-new}
 v^a_{cps} = \left(- v^y_{cps} ,  v^x_{cps} \right) = -\dfrac{1}{4 G_N} \left( \mathfrak{K} \sqrt{y}, 0 \right).\eeq
This vector field also does not match the unit normal \eqref{normal_d1} at the RT surface. So, we need to add a non-trivial vector field $v^a_\ed$ or the harmonic function $\varphi$. Solving Laplace's equation in the background \eqref{d1_new} and using proper boundary conditions, we find
\beq
\varphi = \dfrac{1	}{4 G_N} \left(x-\dfrac{1}{2\sqrt{y_0}} E\left(\dfrac{1}{2}\sec^{-1}\left(\dfrac{y}{y_0}\right)\,\bigg|\, \frac{1}{p_x^2}\right) - \dfrac{\mathfrak{K}}{32} \sqrt{y} -\dfrac{4y_0 y^{3/2}}{(y^2-y_0^2)^{3/2}}\left(\dfrac{x^2}{2} - x \mathfrak{f}(y)\right)\right),
\eeq 
with $\mathfrak{f}(y) = \frac{\sqrt{y^2-y_0^2}}{2 y_0 \sqrt{y}} {_2}F_1\left(\, \frac{1}{4}\, ,\, 1\,;\, \frac{3}{4}\, ;\, \frac{y^2}{y_0^2} \, \right)$ and $E(\mathsf{k}|\mathsf{m})$ is the incomplete elliptic integral of second kind.

\section{Bit threads from PDE}\label{sec:pdebt}
The max-flow formulation of holographic entanglement entropy, seeking a divergenceless vector field $v^a$ with bounded norm $|v| \leq \mathrm{C}$ that maximises flux $\int_A v$, bears a structural resemblance to specific problems involving partial differential equations (PDEs). This appendix aims to outline these connections briefly.

 One relevant connection arises from potential flow theory. If the bit thread field $v^a$ were derivable from a scalar potential $\psi$ (i.e., $v^a= \nabla^a \psi$), the divergenceless condition $\nabla_a v^a = 0$ would translate into the Laplace equation $\nabla^2 \psi = 0$. The problem would then involve solving the Laplace equation subject to the non-linear norm constraint $|\nabla \psi| \leq \mathrm{C}$ and appropriate boundary conditions related to the flux maximisation for region $A$.
 \begin{figure}[htbp]
 \centering
 \includegraphics[width=7.8cm, height=6.2cm]{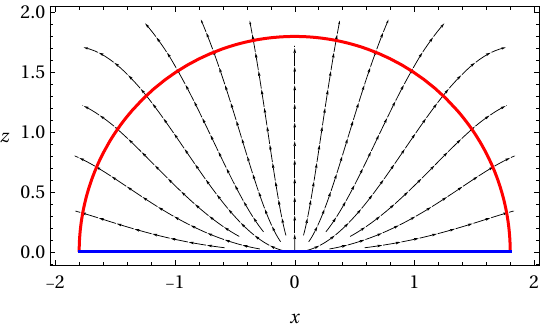}
 \caption{\textit{\label{fig:btPDE}Flow lines for bit threads obtained from solving the PDE: $\nabla_a v_{PDE}^a=0$ and imposing the conditions $v_{PDE}^a|_{RT} =n^a$ and $|v_{PDE}| \le 1$. The blue horizontal line denotes the boundary subregion, and the red semicircle is the corresponding RT surface.}}
 \end{figure}
 
  Here, we construct a flow directly from the PDE: $\nabla_a v_{PDE}^a =0$ and, using the method of characteristics, we find a general solution. Then we impose the conditions  $v_{PDE}^a|_{RT} =n^a/4 G_N$ and $|v_{PDE}| \le 1/4 G_N$. For convenience, we solved the PDE in the Poincar\'{e} AdS$_3$ background \eqref{poincareads} and find a vector field given by,
 \beq
  v_{PDE}^a =\dfrac{1}{4 G_N}\left(\frac{z^2 \left(R \sqrt{R^2-x^2}-R z\right)}{\left(R^2-x^2\right)^{3/2}+x^2}+\frac{z^2}{R} , \frac{x z}{R} \right) .
 \eeq
This is a valid bit thread configuration. Note that this vector is only well defined from the boundary subregion $A$ to the corresponding RT surface $m(A)$. Beyond that, we need to glue some other thread configuration such that $v^a$ remains continuous and differentiable at the RT surface. A similar method was used for `maximally packed flows' in \cite{Agon:2018lwq}. The flow lines of $v_{PDE}^a$ are shown in figure \ref{fig:btPDE}. For simple geometries, this method can also be applied to construct a flow for bit threads.

\section{Coordinate systems of \texorpdfstring{AdS$_3$}{}}\label{sec:coordinates}
Here we will discuss the relation between two different coordinate patches of the AdS spacetime, namely the Poincar\'{e} and Rindler AdS. We will mainly follow the notation of \cite{Casini:2011kv}. The geometry of AdS$_{d+1}$ can be expressed by embedding it to a $\mathbb{R}^{2,d}$ space as
\beq
 - Y_{-1}^2 - Y^2_0 + Y_1^2 + \cdots + Y_d^2 = - L^2 ,
\eeq
with metric
\beq
 ds^2 = - dY_{-1}^2 - dY^2_0 + dY_1^2 + \cdots + dY_d^2  .
\eeq
As discussed in section \ref{sec:btcps}, we are mainly interested in 3-dimensional AdS spacetime. So, the above geometry reduces to AdS$_3$ embedded in $\mathbb{R}^{2,2}$ space as
\beq\label{embedding-hyperboloid}
 - Y_{-1}^2 - Y^2_0 + Y_1^2 + Y_2^2 = - L^2 ,
\eeq
with metric
\beq
 ds^2 = - dY_{-1}^2 - dY^2_0 + dY_1^2 + dY_2^2  .
\eeq
This geometry is connected to the Poincar\'{e} AdS$_3$ via
\beq \label{embedding-poincare}
 Y_{-1} + Y_2 = \dfrac{L^2}{z}, \qquad Y^\m = \dfrac{L}{z} x^\m \quad \mathrm{with} ~\; \m= 0,1.
\eeq
Combining the above equation with the hyperboloid \eqref{embedding-hyperboloid} gives the standard Poincar\'{e} AdS$_3$ metric
\beq\label{Poincare}
ds^2_P = \dfrac{L^2}{z^2} \left(-dt^2 + dx^2 + dz^2  \right),
\eeq
where, $t$ is the time coordinate in Poincar\'{e} patch.

 The Rindler-AdS space is another foliation of AdS$_3$ we are interested in, given by
\beq \label{embedding-rindler}
 Y_{-1} = \rho \cosh{u} , \quad Y_0 = \trho \sinh{\left(\frac{\tau}{L}\right)}, \quad Y_2 = \trho \cosh{\left(\frac{\tau}{L}\right)}, \quad Y_1 = \rho \sinh{u} .
\eeq
Substituting these in \eqref{embedding-hyperboloid} yields $\trho = \sqrt{\rho^2 - L^2}$ and the metric becomes
\beq\label{Rindler}
 ds^2_R = -\left( \dfrac{\trho^2}{L^2} \right) d\tau^2 + \left( \dfrac{\trho^2}{L^2} \right)^{-1} d\rho^2 + \rho^2 du^2 .
\eeq
The above metric gives us a foliation of AdS$_3$ with surfaces of geometry $\mathbb{R \times H}$.\footnote{For AdS$_{d+1}$, the foliation has the geometry of $\mathbb{R \times H}^{d-1}$.} The metric \eqref{Rindler} can be thought of as a topological black hole \cite{Casini:2011kv, Emparan:1999gf} with horizon at $\rho = L$. We are interested in the portion of the Poincar\'{e} AdS$_3$ that is covered by the exterior of the topological black hole. Using the relations \eqref{embedding-poincare} and \eqref{embedding-rindler} we can obtain the following connection between the Poincar\'{e} and Rindler coordinates:
\beq
 Y_{-1} + Y_2 = \dfrac{L^2}{z} = \rho \cosh{u} + \trho \cosh{\left(\frac{\tau}{L}\right)}, \quad  Y_0 =  \dfrac{L}{z} t =\trho \sinh{\left(\frac{\tau}{L}\right)}, \quad Y_1 =  \dfrac{L}{z} x = \rho \sinh{u}.
\eeq
Our main objective is to use the coordinate transformations to find a vector field in terms of one coordinate system from the other on a fixed Cauchy surface. So we will use the above relations with $t=0=\tau$, which is a common Cauchy surface in both coordinate systems. Now, the above relations simplify to
\beq
  Y_{-1} + Y_2 = \dfrac{L^2}{z} = \rho \cosh{u} + \trho , \qquad Y_1=\dfrac{L}{z} x = \rho \sinh{u}.
\eeq
From these two equations, we find the coordinate transformations given by
\begin{subequations}\label{rho,u-z,x}
\beq\label{rho(x,z)}
	\rho = \dfrac{1}{2 z} \sqrt{\left(L^2 +x ^2+z^2\right)^2 - 4L^2x^2},
\eeq
\beq\label{u(x,z)}
	u = \arctanh{\left(\dfrac{2Lx}{L^2+x^2+z^2}\right)}.
\eeq
\end{subequations}
Similarly, the inverse coordinate transformations are
\beq\label{x,z}
z = \dfrac{L^2}{\rho \cosh{u}+\trho}\, , \qquad  x = \dfrac{L \rho \sinh{u}}{\rho \cosh{u}+\trho}.
\eeq
Using these coordinate transformations one can check that the bifurcation surface of \eqref{Rindler} intersects the boundary of the Poincar\'{e} coordinates on a sphere of radius $L$ in the boundary CFT with flat background \cite{Casini:2011kv}. This sphere is the entangling region for which we want to obtain the bit thread configuration.

 The above discussion of coordinate transformations can be extended for a sphere of arbitrary radius $R$ in the CFT \cite{Casini:2011kv}. It implies that the horizon (or more specifically, the bifurcation surface) of the topological black hole moves to a different position, and the intersection with the boundary also changes.  In the Poincar\'{e} coordinates, the change of position of the bifurcation surface corresponds to the different RT surfaces (arcs) corresponding to the boundary entangling region. One can find a family of arcs that are connected by the isometries of the AdS geometry. A boost in the $(Y_{-1}, Y_2)$ plane of the embedding space is one such isometry.  Under this boost, the coordinates transform as
 \beq\label{boost}
  Y'_{-1} = \cosh{\beta} \,Y_{-1} -\sinh{\beta}\, Y_2 , \qquad Y'_2 = \cosh{\beta}\, Y_2 -\sinh{\beta}\, Y_{-1},
 \eeq
while keeping the other coordinates the same. Using these, we again end up with the metric \eqref{Rindler} and choose the Poincar\'{e} coordinates to be the same as in the \eqref{Poincare}. To relate these two coordinates, we need to write $Y_{-1}$ and $Y_2$ in terms of the boost parameter $\beta$ as
\beq\label{embedding-rindler-boost}
 Y_{-1} =\rho \cosh{\beta} \cosh{u} + \trho \sinh{\beta} , \qquad Y_2 = \trho \cosh{\beta} +\rho \sinh{\beta} \cosh{u},
\eeq
on a Cauchy surface (i.e., $\tau = 0 = t$). Now, using the relations \eqref{embedding-poincare}, \eqref{embedding-rindler} and the above ones, we find
\begin{align}
	 Y_{-1} + Y_2 = \dfrac{L^2}{z} &=\rho \cosh{\beta} \cosh{u} + \trho \sinh{\beta} + \trho \cosh{\beta} +\rho \sinh{\beta} \cosh{u}  \nonumber \\
	 &= \left(\rho \cosh{u} + \trho\right) e^\beta, \\ 
	 Y_1 = \dfrac{L}{z} x &= \rho \sinh{u}\,. \nonumber
\end{align}
Similar to the original case, we find the coordinate transformations for the boosted one given by
\begin{subequations}
	\beq\label{rho(x,z)-boost}
	\rho = \dfrac{e^\beta}{2 z} \sqrt{\left(e^{-2\beta}L^2 +x ^2+z^2\right)^2 - 4e^{-2\beta}L^2x^2},
	\eeq
	\beq\label{u(x,z)-boost}
	u = \arctanh{\left(\dfrac{2e^{-\beta}Lx}{e^{-2\beta}L^2+x^2+z^2}\right)}.
	\eeq
\end{subequations}
The inverse transformations are
\beq\label{x,z-boost}
z = \dfrac{e^{-\beta}L^2}{\rho \cosh{u}+\trho}\, , \qquad  x = \dfrac{e^{-\beta}L \rho \sinh{u}}{\rho \cosh{u}+\trho}.
\eeq

	\bibliographystyle{unsrt}
	\bibliography{ref.bib}
\end{document}